\documentclass[twoside,twocolumn,english,pra,aps,superscriptaddress]{revtex4-1}
\usepackage[T1]{fontenc}
\usepackage{color}
\usepackage{babel}

\usepackage{bbm}
\usepackage{textcomp}

\usepackage{amssymb,amsfonts,mathrsfs}
\usepackage{amsthm}
\usepackage{amsmath}
\usepackage{graphicx}

\usepackage{ulem}

\usepackage[unicode=true,pdfusetitle,
 bookmarks=true,bookmarksnumbered=false,bookmarksopen=false,
 breaklinks=false,pdfborder={0 0 0},backref=false,colorlinks=true]
 {hyperref}
\usepackage{breakurl}
\usepackage{tikz}
\usetikzlibrary{positioning, shapes.geometric}

\makeatletter
\theoremstyle{plain}

\makeatother

\begin{document}

\preprint{This line only printed with preprint option}

\title{Exact correlation functions for dual-unitary quantum circuits with exceptional points}

\author{Xi-Dan Hu}
\affiliation {Key Laboratory of Atomic and Subatomic Structure and Quantum Control (Ministry of Education), Guangdong Basic Research Center of Excellence for Structure and Fundamental Interactions of Matter, School of Physics, South China Normal University, Guangzhou 510006, China}

\author{Dan-Bo Zhang}
\email{dbzhang@m.scnu.edu.cn}
\affiliation {Key Laboratory of Atomic and Subatomic Structure and Quantum Control (Ministry of Education), Guangdong Basic Research Center of Excellence for Structure and Fundamental Interactions of Matter, School of Physics, South China Normal University, Guangzhou 510006, China}

\affiliation{Guangdong Provincial Key Laboratory of Quantum Engineering and Quantum Materials,  Guangdong-Hong Kong Joint Laboratory of Quantum Matter, and Frontier Research Institute for Physics,\\  South China Normal University, Guangzhou 510006, China}

\date{\today}

\begin{abstract}
Dual-unitary quantum circuits can provide analytic spatiotemporal correlation functions of local operators from transfer matrices, enriching our understanding of quantum dynamics with exact solutions. Nevertheless, a full understanding is still lacking as the case of a non-diagonalizable transfer matrix with exceptional points has hardly been explored.
In this paper, we give an inverse approach for constructing dual-unitary quantum circuits with exceptional points in the transfer matrices, by establishing relations between transfer matrices and local unitaries. As a consequence of the coalesce of eigenvectors, the correlation functions exhibit a polynomial modified exponential decay, which is significantly different from pure exponential decay, especially at early stages. Moreover, we point out that the Hamiltonian evolution of a kicked XXZ spin chain can be approximately mapped to a dual-unitary circuit with exceptional points by Trotter decomposition. Finally, we investigate the dynamics approaching and at exceptional points, showing that behaviors of correlation functions are distinct by $Z$ transformation.
\end{abstract}

\maketitle

\section{Introduction}
Spatiotemporal correlation functions provide a basic and useful understanding of many-body dynamical properties, which can be used to access characteristics of the physical system, e.g., matter transporting and information spreading. Moreover, while experimentally observable~\cite{ImmanuelBloch2008,GeraldDMahan2011,MarcCheneau2012}, spatiotemporal correlation functions are notoriously hard, even unable to be accessed in most interacting many-body systems due to the exponential wall of Hilbert space dimension (for numerical method)~\cite{WKohn1999} or unsolvable problem (for analytical method).
There is an absence of general methods to exactly solve in a large site system with long-time evolution.
Although tensor network (TN)~\cite{ShiJuRan2020} provides a great numerical method to calculate more sites rather than the exact diagonalization method, the long-time evolution of TN is limited by time entanglement barrier (TEB)~\cite{MichaelSonner2021,ErikaYe2021}, with a few exceptional for specified systems~\cite{AlessioLerose2023}. Since numerical methods typically introduce errors, it is very attractive to find a method that can exactly solve the spatiotemporal correlation functions.


On the other hand, the dual-unitary quantum circuit provides exact solvable spatiotemporal correlation functions~\cite{BrunoBertini2019}. The dual-unitary circuit is composed of a series of local evolution operators satisfying dual-unitary, which are unitary both in time and space direction~\cite{BrunoBertini2019,CheryneJonay2021,RichardMMilbradt2023,XieHangYu2023}.
The dual-unitary circuits allow investigations of quantum dynamics of both integrable systems and chaotic systems. Remarkably, it provides a perfect platform to study the quantum chaos exactly~\cite{BrunoBertini2020a,TomazProsen2021,PieterWClaeys2022a}, by calculating indicators of quantum chaos exactly, such as out-of-time-order correlators~\cite{PieterWClaeys2020,PieterWClaeys2022c,JasBensa2022,MichaelARampp2023} and operator spreading~\cite{TamasGombor2022,MichaelARampp2023,GraceMSommers2023}.
Moreover, dual-unitary quantum circuits have been widely used in the study of many-body dynamics, such as ergodicity~\cite{BrunoBertini2019,LorenzoPiroli2020,PieterWClaeys2021,SAravinda2021,MartonBorsi2022,FelixFritzsch2023}, entanglement properties~\cite{BrunoBertini2020a,BrunoBertini2020b,SuhailAhmadRather2020,IsaacReid2021,MatteoIppoliti2021,TianciZhou2022,MatteoIppoliti2022,AZabalo2022,PieterWClaeys2022b,FelixFritzsch2023,BrunoBertini2023,AlessandroFoligno2023a,MichaelARampp2023,AlessandroFoligno2023b,MartonMestyan2024}, quantum many-body dynamics~\cite{BorisGutkin2020,FelixFritzsch2021,AlessioLerose2021,PavelKos2021,BalazsPozsgay2021,BrunoBertini2022,MatteoIppoliti2023,DominikHahn2023,PieterWClaeys2023,LeonardLogaric2024}, and quantum information scrambling~\cite{BrunoBertini2020c,FelixFritzsch2022,GraceMSommers2023,NeilDowling2023,AlessandroFoligno2024}. The classification of dual-unitary circuits relies on the transfer matrix, which can be derived solely from the local dual-unitary gates and fully capture the information of dynamics. In this regard, the transfer matrix plays a central role. While the classification of dual-unitary circuits for qubits ($d=2$) has been given by analyzing the transfer matrix~\cite{BrunoBertini2019}, the case of a non-diagonalizable transfer matrix with exceptional points~(EPs), as well as the physical consequence, has not been explicitly explored.

In this paper, we give explicit constructions of dual-unitary quantum circuits whose transfer matrices have EPs and investigate their physical consequences. The construction follows an inverse method, which starts from a transfer matrix with a Jordan block and then derives the local dual-unitary gates, by leveraging constraints of the relations between transfer matrices and the local dual-unitary gates. We find that the transfer matrix with EPs can introduce a polynomial modification on the exponential decay spatiotemporal correlation functions, which stems from the coalesce of eigenvectors. Specifically, we consider both $2\times2$ and $3\times3$ Jordan blocks cases of the transfer matrix for qubits.
Correspondingly, we propose lattice models of kicked spin XXZ chains whose Hamiltonian evolution at infinite temperature can be approximated by the dual-unitary circuits, due to both of them having similar behaviors of the spatiotemporal correlation functions.
Finally, we investigate the dynamics of dual-unitary circuits approaching and at EPs.
We find that one can distinguish three different behaviors of the correlation function nearing and at EPs by $Z$ transformation.

The paper is organized as follows. In Sec.~\ref{s2}, we review basic concepts and techniques for dual-unitary quantum circuits.  In Sec.~\ref{s3} we give constructions of dual-unitary quantum circuits whose transfer matrices have EPs.  In Sec.~\ref{s4}, we study distinct dynamical properties for dual-unitary circuits approaching and at EPs. Finally, we give conclusions at Sec.~\ref{s5}.

\section{Exactly solvable spatiotemporal correlation function in dual-unitary quantum circuit}
\label{s2}
In this section, we briefly review the framework of dual-unitary quantum circuits, and how to obtain the exactly solvable spatiotemporal correlation function.

The dual-unitary circuit consists of a series of two-qubit dual-unitary local gates $U$. The dual-unitary properties of $U$ can be written as,
\begin{equation}
\begin{split}
U^\dag U=&UU^\dag=\mathbbm{1}\\
\tilde{U}^\dag \tilde{U}=&\tilde{U}\tilde{U}^\dag=\mathbbm{1},\\
\end{split}
\end{equation}
where $\tilde{U}$ is the dual local gate of $U$. The dual gate $\tilde{U}$ is related to $U$ by the following reshuffling~\cite{BrunoBertini2019},
\begin{equation}
\langle k|\otimes \langle l|\tilde{U}|i\rangle\otimes|j\rangle=\langle j|\otimes \langle l|U|i\rangle\otimes|k\rangle.
\end{equation}
Corresponding to the reshuffling of indexes, the roles of time and space for $\tilde{U}$ and $U$ in the evolution are swapped in a quantum circuit.

It is convenient to adopt a diagrammatic representation.
A local gate can be described by a square and a series of legs, where each leg corresponds to a Hilbert space of one qubit.
Moreover, a single qubit gate represents a bullet $\bullet$. Then the two qubit local gate and its Hermitian conjugate are represented as,

\begin{equation}
\begin{tikzpicture}
\node at(-1,0) {$U=$};
\filldraw[draw,fill=red!100,scale=0.25,shift={(0,0)},line width =1pt] (-1,-1) rectangle (1,1);
\draw[scale=0.25,shift={(0,0)},line width =1pt] (-1,-1)--(-2,-2);
\draw[scale=0.25,shift={(0,0)},line width =1pt] (-1,1)--(-2,2);
\draw [scale=0.25,shift={(0,0)},line width =1pt] (1,-1)--(2,-2);
\draw [scale=0.25,shift={(0,0)},line width =1pt] (1,1)--(2,2);

\node at(0.65,0.65) {$l$};
\node at(-0.65,0.65) {$k$};
\node at(0.65,-0.65) {$j$};
\node at(-0.65,-0.65) {$i$};

\node at(0.8,0) {,};

\node at(2.5,0) {$U^{\dag}=$};
\filldraw[draw,fill=blue!100,scale=0.25,shift={(14,0)},line width =1pt] (-1,-1) rectangle (1,1);
\draw[scale=0.25,shift={(14,0)},line width =1pt] (-1,-1)--(-2,-2);
\draw[scale=0.25,shift={(14,0)},line width =1pt] (-1,1)--(-2,2);
\draw [scale=0.25,shift={(14,0)},line width =1pt] (1,-1)--(2,-2);
\draw [scale=0.25,shift={(14,0)},line width =1pt] (1,1)--(2,2);

\node at(0.65+3.5,0.65) {$j$};
\node at(-0.65+3.5,0.65) {$i$};
\node at(0.65+3.5,-0.65) {$l$};
\node at(-0.65+3.5,-0.65) {$k$};

\node at(4.2,0) {.};
\end{tikzpicture}
\end{equation}
Now the unitarity of $U$ and $\tilde{U}$ can be viewed as,
\begin{equation}
\begin{tikzpicture}
\node at(-2.4,0) {$UU^\dag=U^\dag U=\mathbbm{1}$  \quad$\Rightarrow $};
\filldraw[draw,fill=red!100,scale=0.15,shift={(0,3)},line width =0.75pt] (-1,-1) rectangle (1,1);
\draw[scale=0.15,shift={(0,3)},line width =1pt] (-1,1)--(-2,2);
\draw [scale=0.15,shift={(0,3)},line width =1pt] (1,1)--(2,2);

\draw [scale=0.15,shift={(0,3)},line width =1pt] (-1,-1)--(-1,-5);
\draw [scale=0.15,shift={(0,3)},line width =1pt] (1,-1)--(1,-5);

\filldraw[draw,fill=blue!100,scale=0.15,shift={(0,-3)},line width =0.75pt] (-1,-1) rectangle (1,1);
\draw[scale=0.15,shift={(0,-3)},line width =1pt] (-1,-1)--(-2,-2);
\draw [scale=0.15,shift={(0,-3)},line width =1pt] (1,-1)--(2,-2);

\node at(0.8,0) {$=$};

\filldraw[draw,fill=blue!100,scale=0.15,shift={(11,3)},line width =0.75pt] (-1,-1) rectangle (1,1);
\draw[scale=0.15,shift={(11,3)},line width =1pt] (-1,1)--(-2,2);
\draw [scale=0.15,shift={(11,3)},line width =1pt] (1,1)--(2,2);

\draw [scale=0.15,shift={(11,3)},line width =1pt] (-1,-1)--(-1,-5);
\draw [scale=0.15,shift={(11,3)},line width =1pt] (1,-1)--(1,-5);

\filldraw[draw,fill=red!100,scale=0.15,shift={(11,-3)},line width =0.75pt] (-1,-1) rectangle (1,1);
\draw[scale=0.15,shift={(11,-3)},line width =1pt] (-1,-1)--(-2,-2);
\draw [scale=0.15,shift={(11,-3)},line width =1pt] (1,-1)--(2,-2);

\node at(2.5,0) {$=$};

\draw[scale=0.15,shift={(22,3)},line width =1pt] (-1,1)--(-2,2);
\draw [scale=0.15,shift={(22,3)},line width =1pt] (1,1)--(2,2);
\draw[scale=0.15,shift={(22,3)},line width =1pt] (-1,1+0.05)--(-1,-7-0.05);
\draw [scale=0.15,shift={(22,3)},line width =1pt] (1,1+0.05)--(1,-7-0.05);
\draw[scale=0.15,shift={(22,-3)},line width =1pt] (-1,-1)--(-2,-2);
\draw [scale=0.15,shift={(22,-3)},line width =1pt] (1,-1)--(2,-2);

\node at(3.7,0) {,};
\end{tikzpicture}
\end{equation}
and,
\begin{equation}
\begin{tikzpicture}
\node at(-2.5,0) {$\tilde{U}\tilde{U}^\dag=\tilde{U}^\dag \tilde{U}=\mathbbm{1}$  \quad$\Rightarrow $};
\filldraw[draw,fill=red!100,scale=0.15,shift={(0,3)},line width =0.75pt] (-1,-1) rectangle (1,1);
\draw [scale=0.15,shift={(0,3)},line width =1pt] (1,-1)--(2,-2);
\draw [scale=0.15,shift={(0,3)},line width =1pt] (1,1)--(2,2);

\draw [scale=0.15,shift={(0,3)},line width =1pt] (-1,-1)--(-1,-5);
\draw [scale=0.15,shift={(0,3)},line width =1pt] (-1,1)--(-3,1)--(-3,-7)--(-1,-7);

\filldraw[draw,fill=blue!100,scale=0.15,shift={(0,-3)},line width =0.75pt] (-1,-1) rectangle (1,1);
\draw [scale=0.15,shift={(0,-3)},line width =1pt] (1,-1)--(2,-2);
\draw [scale=0.15,shift={(0,-3)},line width =1pt] (1,1)--(2,2);

\node at(0.6,0) {$=$};

\draw [scale=0.15,shift={(9,3)},line width =1pt] (1,-1)--(2,-2);
\draw [scale=0.15,shift={(9,3)},line width =1pt] (1,1)--(2,2);

\draw [scale=0.15,shift={(9,3)},line width =1pt] (1+0.1,-1)--(-1,-1);
\draw [scale=0.15,shift={(9,3)},line width =1pt] (1+0.1,1)--(-1,1);

\draw [scale=0.15,shift={(9,3)},line width =1pt] (-1,-1+0.1)--(-1,-5-0.1);
\draw [scale=0.15,shift={(9,3)},line width =1pt] (-1,1)--(-3,1)--(-3,-7)--(-1,-7);

\draw [scale=0.15,shift={(9,-3)},line width =1pt] (1+0.1,-1)--(-1,-1);
\draw [scale=0.15,shift={(9,-3)},line width =1pt] (1+0.1,1)--(-1,1);

\draw [scale=0.15,shift={(9,-3)},line width =1pt] (1,-1)--(2,-2);
\draw [scale=0.15,shift={(9,-3)},line width =1pt] (1,1)--(2,2);

\node at(1.75,0) {,};

\filldraw[draw,fill=red!100,scale=0.15,shift={(16,3)},line width =0.75pt] (-1,-1) rectangle (1,1);
\draw [scale=0.15,shift={(16,3)},line width =1pt] (-1,-1)--(-2,-2);
\draw [scale=0.15,shift={(16,3)},line width =1pt] (-1,1)--(-2,2);

\draw [scale=0.15,shift={(16,3)},line width =1pt] (1,-1)--(1,-5);
\draw [scale=0.15,shift={(16,3)},line width =1pt] (1,1)--(3,1)--(3,-7)--(1,-7);

\filldraw[draw,fill=blue!100,scale=0.15,shift={(16,-3)},line width =0.75pt] (-1,-1) rectangle (1,1);
\draw [scale=0.15,shift={(16,-3)},line width =1pt] (-1,-1)--(-2,-2);
\draw [scale=0.15,shift={(16,-3)},line width =1pt] (-1,1)--(-2,2);

\node at(3.1,0) {$=$};

\draw [scale=0.15,shift={(24,3)},line width =1pt] (-1,-1)--(-2,-2);
\draw [scale=0.15,shift={(24,3)},line width =1pt] (-1,1)--(-2,2);

\draw [scale=0.15,shift={(24,3)},line width =1pt] (-1-0.1,-1)--(1,-1);
\draw [scale=0.15,shift={(24,3)},line width =1pt] (-1-0.1,1)--(1,1);

\draw [scale=0.15,shift={(24,3)},line width =1pt] (1,-1+0.1)--(1,-5-0.1);
\draw [scale=0.15,shift={(24,3)},line width =1pt] (1,1)--(3,1)--(3,-7)--(1,-7);

\draw [scale=0.15,shift={(24,-3)},line width =1pt] (-1-0.1,-1)--(1,-1);
\draw [scale=0.15,shift={(24,-3)},line width =1pt] (-1-0.1,1)--(1,1);

\draw [scale=0.15,shift={(24,-3)},line width =1pt] (-1,-1)--(-2,-2);
\draw [scale=0.15,shift={(24,-3)},line width =1pt] (-1,1)--(-2,2);

\node at(4.2,0) {.};
\end{tikzpicture}
\end{equation}
Those two-qubit gates can construct a quantum circuit. The unitary transformation for each two layers of the quantum circuit is denoted as,
\begin{equation}
\begin{tikzpicture}
\node at(-0.9,0.25) {\fontsize{12pt}{\baselineskip}\selectfont $\mathbbm{U}\Rightarrow$};

\node at(-0.12*2,-0.12*5) {-$\frac{5}{2}$};
\node at(0.12*2,-0.12*5) {\fontsize{7pt}{\baselineskip}\selectfont {-$2$}};
\node at(0.12*6,-0.12*5) {-$\frac{3}{2}$};
\node at(0.12*10,-0.12*5) {\fontsize{7pt}{\baselineskip}\selectfont {-$1$}};
\node at(0.12*14,-0.12*5) {-$\frac{1}{2}$};
\node at(0.12*18,-0.12*5) {\fontsize{7pt}{\baselineskip}\selectfont {$0$}};
\node at(0.12*22,-0.12*5) {$\frac{1}{2}$};
\node at(0.12*26,-0.12*5) {\fontsize{7pt}{\baselineskip}\selectfont {$1$}};
\node at(0.12*30,-0.12*5) {$\frac{3}{2}$};
\node at(0.12*34,-0.12*5) {\fontsize{7pt}{\baselineskip}\selectfont {$2$}};
\node at(0.12*38,-0.12*5) {$\frac{5}{2}$};
\node at(0.12*42,-0.12*5) {\fontsize{7pt}{\baselineskip}\selectfont {$3$}};
\node at(0.12*46,-0.12*5) {-$\frac{5}{2}$};
\node at(0.12*18,-0.12*8) {\fontsize{7pt}{\baselineskip}\selectfont {$site$}};

\node at(0.12*49,-0.12*2) {\fontsize{7pt}{\baselineskip}\selectfont {$1$}};
\node at(0.12*49,0.12*2) {$\frac{1}{2}$};
\node at(0.12*49,0.12*6) {\fontsize{7pt}{\baselineskip}\selectfont {$0$}};
\node at(0.12*52,0.12*2) {\fontsize{7pt}{\baselineskip}\selectfont {$t$}};

\node at(0.12*54,0.12*1) {\fontsize{7pt}{\baselineskip}\selectfont {$.$}};

\filldraw[draw,fill=red!100,scale=0.12,shift={(0,0)},line width =0.5pt] (-1,-1) rectangle (1,1);
\draw [scale=0.12,shift={(0,0)},line width =0.75pt] (1,1)--(2,2);
\draw [scale=0.12,shift={(0,0)},line width =0.75pt] (-1,1)--(-2,2);
\draw [scale=0.12,shift={(0,0)},line width =0.75pt] (1,-1)--(2,-2);
\draw [scale=0.12,shift={(0,0)},line width =0.75pt] (-1,-1)--(-2,-2);

\filldraw[draw,fill=red!100,scale=0.12,shift={(8,0)},line width =0.5pt] (-1,-1) rectangle (1,1);
\draw [scale=0.12,shift={(8,0)},line width =0.75pt] (1,1)--(2,2);
\draw [scale=0.12,shift={(8,0)},line width =0.75pt] (-1,1)--(-2,2);
\draw [scale=0.12,shift={(8,0)},line width =0.75pt] (1,-1)--(2,-2);
\draw [scale=0.12,shift={(8,0)},line width =0.75pt] (-1,-1)--(-2,-2);

\filldraw[draw,fill=red!100,scale=0.12,shift={(16,0)},line width =0.5pt] (-1,-1) rectangle (1,1);
\draw [scale=0.12,shift={(16,0)},line width =0.75pt] (1,1)--(2,2);
\draw [scale=0.12,shift={(16,0)},line width =0.75pt] (-1,1)--(-2,2);
\draw [scale=0.12,shift={(16,0)},line width =0.75pt] (1,-1)--(2,-2);
\draw [scale=0.12,shift={(16,0)},line width =0.75pt] (-1,-1)--(-2,-2);

\filldraw[draw,fill=red!100,scale=0.12,shift={(24,0)},line width =0.5pt] (-1,-1) rectangle (1,1);
\draw [scale=0.12,shift={(24,0)},line width =0.75pt] (1,1)--(2,2);
\draw [scale=0.12,shift={(24,0)},line width =0.75pt] (-1,1)--(-2,2);
\draw [scale=0.12,shift={(24,0)},line width =0.75pt] (1,-1)--(2,-2);
\draw [scale=0.12,shift={(24,0)},line width =0.75pt] (-1,-1)--(-2,-2);

\filldraw[draw,fill=red!100,scale=0.12,shift={(32,0)},line width =0.5pt] (-1,-1) rectangle (1,1);
\draw [scale=0.12,shift={(32,0)},line width =0.75pt] (1,1)--(2,2);
\draw [scale=0.12,shift={(32,0)},line width =0.75pt] (-1,1)--(-2,2);
\draw [scale=0.12,shift={(32,0)},line width =0.75pt] (1,-1)--(2,-2);
\draw [scale=0.12,shift={(32,0)},line width =0.75pt] (-1,-1)--(-2,-2);

\filldraw[draw,fill=red!100,scale=0.12,shift={(40,0)},line width =0.5pt] (-1,-1) rectangle (1,1);
\draw [scale=0.12,shift={(40,0)},line width =0.75pt] (1,1)--(2,2);
\draw [scale=0.12,shift={(40,0)},line width =0.75pt] (-1,1)--(-2,2);
\draw [scale=0.12,shift={(40,0)},line width =0.75pt] (1,-1)--(2,-2);
\draw [scale=0.12,shift={(40,0)},line width =0.75pt] (-1,-1)--(-2,-2);

\filldraw[draw,fill=red!100,scale=0.12,shift={(4,4)},line width =0.5pt] (-1,-1) rectangle (1,1);
\draw [scale=0.12,shift={(4,4)},line width =0.75pt] (1,1)--(2,2);
\draw [scale=0.12,shift={(4,4)},line width =0.75pt] (-1,1)--(-2,2);
\draw [scale=0.12,shift={(4,4)},line width =0.75pt] (1,-1)--(2,-2);
\draw [scale=0.12,shift={(4,4)},line width =0.75pt] (-1,-1)--(-2,-2);

\filldraw[draw,fill=red!100,scale=0.12,shift={(12,4)},line width =0.5pt] (-1,-1) rectangle (1,1);
\draw [scale=0.12,shift={(12,4)},line width =0.75pt] (1,1)--(2,2);
\draw [scale=0.12,shift={(12,4)},line width =0.75pt] (-1,1)--(-2,2);
\draw [scale=0.12,shift={(12,4)},line width =0.75pt] (1,-1)--(2,-2);
\draw [scale=0.12,shift={(12,4)},line width =0.75pt] (-1,-1)--(-2,-2);

\filldraw[draw,fill=red!100,scale=0.12,shift={(20,4)},line width =0.5pt] (-1,-1) rectangle (1,1);
\draw [scale=0.12,shift={(20,4)},line width =0.75pt] (1,1)--(2,2);
\draw [scale=0.12,shift={(20,4)},line width =0.75pt] (-1,1)--(-2,2);
\draw [scale=0.12,shift={(20,4)},line width =0.75pt] (1,-1)--(2,-2);
\draw [scale=0.12,shift={(20,4)},line width =0.75pt] (-1,-1)--(-2,-2);

\filldraw[draw,fill=red!100,scale=0.12,shift={(28,4)},line width =0.5pt] (-1,-1) rectangle (1,1);
\draw [scale=0.12,shift={(28,4)},line width =0.75pt] (1,1)--(2,2);
\draw [scale=0.12,shift={(28,4)},line width =0.75pt] (-1,1)--(-2,2);
\draw [scale=0.12,shift={(28,4)},line width =0.75pt] (1,-1)--(2,-2);
\draw [scale=0.12,shift={(28,4)},line width =0.75pt] (-1,-1)--(-2,-2);

\filldraw[draw,fill=red!100,scale=0.12,shift={(36,4)},line width =0.5pt] (-1,-1) rectangle (1,1);
\draw [scale=0.12,shift={(36,4)},line width =0.75pt] (1,1)--(2,2);
\draw [scale=0.12,shift={(36,4)},line width =0.75pt] (-1,1)--(-2,2);
\draw [scale=0.12,shift={(36,4)},line width =0.75pt] (1,-1)--(2,-2);
\draw [scale=0.12,shift={(36,4)},line width =0.75pt] (-1,-1)--(-2,-2);

\filldraw[draw,fill=red!100,scale=0.12,shift={(44,4)},line width =0.5pt] (-1,-1) rectangle (1,1);
\draw [scale=0.12,shift={(44,4)},line width =0.75pt] (1,1)--(2,2);
\draw [scale=0.12,shift={(44,4)},line width =0.75pt] (-1,1)--(-2,2);
\draw [scale=0.12,shift={(44,4)},line width =0.75pt] (1,-1)--(2,-2);
\draw [scale=0.12,shift={(44,4)},line width =0.75pt] (-1,-1)--(-2,-2);

\end{tikzpicture}
\end{equation}
The dynamical properties of the quantum circuit can be captured by the dynamical correlation functions. To be exactly solvable, the initial state of the dual-unitary quantum circuit should be specific, which can be an infinite temperature state~(a maximally mixed state) or some specific matrix product state~\cite{LorenzoPiroli2020,FelixFritzsch2023,PieterWClaeys2022}. We focus on the dynamical correlation functions of local operators in the infinite temperature state, which can be expressed as~\cite{BrunoBertini2019},
\begin{equation}\label{DCF}
D^{\alpha,\beta}(x,y,t)=\frac{1}{2^{2L}}\text{tr}[\sigma_x^\alpha\mathbbm{U}^{-t}\sigma_y^\beta\mathbbm{U}^t],
\end{equation}
where $x$, $y$ $\in$ $\frac{1}{2}\mathbbm{Z}_{2L}$ is the site label, $t\in \mathbbm{N}$ and $\sigma_n^{\alpha}$ ($\alpha=x,y,z$) denotes the Pauli matrix at the site $n$. Here, $\sigma^0=\mathbbm{1}_2$ denotes a $2\times2$ identity matrix. Graphically,
the equation~\eqref{DCF} can be represented as,
\begin{equation}
\begin{tikzpicture}
\node at(-0.9,2.7) {\fontsize{12pt}{\baselineskip}\selectfont $\frac{1}{2^{2L}}$};

\node at(-0.12*2,-0.12*5) {-$\frac{5}{2}$};
\node at(0.12*2,-0.12*5) {\fontsize{7pt}{\baselineskip}\selectfont {-$2$}};
\node at(0.12*6,-0.12*5) {-$\frac{3}{2}$};
\node at(0.12*10,-0.12*5) {\fontsize{7pt}{\baselineskip}\selectfont {-$1$}};
\node at(0.12*14,-0.12*5) {-$\frac{1}{2}$};
\node at(0.12*18,-0.12*5) {\fontsize{7pt}{\baselineskip}\selectfont {$0$}};
\node at(0.12*22,-0.12*5) {$\frac{1}{2}$};
\node at(0.12*26,-0.12*5) {\fontsize{7pt}{\baselineskip}\selectfont {$1$}};
\node at(0.12*30,-0.12*5) {$\frac{3}{2}$};
\node at(0.12*34,-0.12*5) {\fontsize{7pt}{\baselineskip}\selectfont {$2$}};
\node at(0.12*38,-0.12*5) {$\frac{5}{2}$};
\node at(0.12*42,-0.12*5) {\fontsize{7pt}{\baselineskip}\selectfont {$3$}};
\node at(0.12*46,-0.12*5) {-$\frac{5}{2}$};
\node at(0.12*18,-0.12*8) {\fontsize{7pt}{\baselineskip}\selectfont {$site$}};

\node at(0.12*49,0.12*2) {$\frac{5}{2}$};
\node at(0.12*49,0.12*6) {\fontsize{7pt}{\baselineskip}\selectfont {$2$}};
\node at(0.12*49,0.12*10) {$\frac{3}{2}$};
\node at(0.12*49,0.12*14) {\fontsize{7pt}{\baselineskip}\selectfont {$1$}};
\node at(0.12*49,0.12*18) {$\frac{1}{2}$};
\node at(0.12*49,0.12*22) {\fontsize{7pt}{\baselineskip}\selectfont {$0$}};
\node at(0.12*52,0.12*22) {\fontsize{7pt}{\baselineskip}\selectfont {$t$}};
\node at(0.12*49,0.12*26) {-$\frac{1}{2}$};
\node at(0.12*49,0.12*30) {\fontsize{7pt}{\baselineskip}\selectfont {-$1$}};
\node at(0.12*49,0.12*34) {-$\frac{3}{2}$};
\node at(0.12*49,0.12*38) {\fontsize{7pt}{\baselineskip}\selectfont {-$2$}};
\node at(0.12*49,0.12*42) {-$\frac{5}{2}$};

\node at(0.12*54,0.12*21) {\fontsize{7pt}{\baselineskip}\selectfont {$,$}};

\filldraw[draw,fill=red!100,scale=0.12,shift={(0,0)},line width =0.5pt] (-1,-1) rectangle (1,1);
\draw [scale=0.12,shift={(0,0)},line width =0.75pt] (1,1)--(2,2);
\draw [scale=0.12,shift={(0,0)},line width =0.75pt] (-1,1)--(-2,2);
\draw [scale=0.12,shift={(0,0)},line width =0.75pt] (1,-1)--(2,-2);
\draw [scale=0.12,shift={(0,0)},line width =0.75pt] (-1,-1)--(-2,-2);

\filldraw[draw,fill=red!100,scale=0.12,shift={(8,0)},line width =0.5pt] (-1,-1) rectangle (1,1);
\draw [scale=0.12,shift={(8,0)},line width =0.75pt] (1,1)--(2,2);
\draw [scale=0.12,shift={(8,0)},line width =0.75pt] (-1,1)--(-2,2);
\draw [scale=0.12,shift={(8,0)},line width =0.75pt] (1,-1)--(2,-2);
\draw [scale=0.12,shift={(8,0)},line width =0.75pt] (-1,-1)--(-2,-2);

\filldraw[draw,fill=red!100,scale=0.12,shift={(16,0)},line width =0.5pt] (-1,-1) rectangle (1,1);
\draw [scale=0.12,shift={(16,0)},line width =0.75pt] (1,1)--(2,2);
\draw [scale=0.12,shift={(16,0)},line width =0.75pt] (-1,1)--(-2,2);
\draw [scale=0.12,shift={(16,0)},line width =0.75pt] (1,-1)--(2,-2);
\draw [scale=0.12,shift={(16,0)},line width =0.75pt] (-1,-1)--(-2,-2);

\filldraw[draw,fill=red!100,scale=0.12,shift={(24,0)},line width =0.5pt] (-1,-1) rectangle (1,1);
\draw [scale=0.12,shift={(24,0)},line width =0.75pt] (1,1)--(2,2);
\draw [scale=0.12,shift={(24,0)},line width =0.75pt] (-1,1)--(-2,2);
\draw [scale=0.12,shift={(24,0)},line width =0.75pt] (1,-1)--(2,-2);
\draw [scale=0.12,shift={(24,0)},line width =0.75pt] (-1,-1)--(-2,-2);

\filldraw[draw,fill=red!100,scale=0.12,shift={(32,0)},line width =0.5pt] (-1,-1) rectangle (1,1);
\draw [scale=0.12,shift={(32,0)},line width =0.75pt] (1,1)--(2,2);
\draw [scale=0.12,shift={(32,0)},line width =0.75pt] (-1,1)--(-2,2);
\draw [scale=0.12,shift={(32,0)},line width =0.75pt] (1,-1)--(2,-2);
\draw [scale=0.12,shift={(32,0)},line width =0.75pt] (-1,-1)--(-2,-2);

\filldraw[draw,fill=red!100,scale=0.12,shift={(40,0)},line width =0.5pt] (-1,-1) rectangle (1,1);
\draw [scale=0.12,shift={(40,0)},line width =0.75pt] (1,1)--(2,2);
\draw [scale=0.12,shift={(40,0)},line width =0.75pt] (-1,1)--(-2,2);
\draw [scale=0.12,shift={(40,0)},line width =0.75pt] (1,-1)--(2,-2);
\draw [scale=0.12,shift={(40,0)},line width =0.75pt] (-1,-1)--(-2,-2);

\filldraw[draw,fill=red!100,scale=0.12,shift={(4,4)},line width =0.5pt] (-1,-1) rectangle (1,1);
\draw [scale=0.12,shift={(4,4)},line width =0.75pt] (1,1)--(2,2);
\draw [scale=0.12,shift={(4,4)},line width =0.75pt] (-1,1)--(-2,2);
\draw [scale=0.12,shift={(4,4)},line width =0.75pt] (1,-1)--(2,-2);
\draw [scale=0.12,shift={(4,4)},line width =0.75pt] (-1,-1)--(-2,-2);

\filldraw[draw,fill=red!100,scale=0.12,shift={(12,4)},line width =0.5pt] (-1,-1) rectangle (1,1);
\draw [scale=0.12,shift={(12,4)},line width =0.75pt] (1,1)--(2,2);
\draw [scale=0.12,shift={(12,4)},line width =0.75pt] (-1,1)--(-2,2);
\draw [scale=0.12,shift={(12,4)},line width =0.75pt] (1,-1)--(2,-2);
\draw [scale=0.12,shift={(12,4)},line width =0.75pt] (-1,-1)--(-2,-2);

\filldraw[draw,fill=red!100,scale=0.12,shift={(20,4)},line width =0.5pt] (-1,-1) rectangle (1,1);
\draw [scale=0.12,shift={(20,4)},line width =0.75pt] (1,1)--(2,2);
\draw [scale=0.12,shift={(20,4)},line width =0.75pt] (-1,1)--(-2,2);
\draw [scale=0.12,shift={(20,4)},line width =0.75pt] (1,-1)--(2,-2);
\draw [scale=0.12,shift={(20,4)},line width =0.75pt] (-1,-1)--(-2,-2);

\filldraw[draw,fill=red!100,scale=0.12,shift={(28,4)},line width =0.5pt] (-1,-1) rectangle (1,1);
\draw [scale=0.12,shift={(28,4)},line width =0.75pt] (1,1)--(2,2);
\draw [scale=0.12,shift={(28,4)},line width =0.75pt] (-1,1)--(-2,2);
\draw [scale=0.12,shift={(28,4)},line width =0.75pt] (1,-1)--(2,-2);
\draw [scale=0.12,shift={(28,4)},line width =0.75pt] (-1,-1)--(-2,-2);

\filldraw[draw,fill=red!100,scale=0.12,shift={(36,4)},line width =0.5pt] (-1,-1) rectangle (1,1);
\draw [scale=0.12,shift={(36,4)},line width =0.75pt] (1,1)--(2,2);
\draw [scale=0.12,shift={(36,4)},line width =0.75pt] (-1,1)--(-2,2);
\draw [scale=0.12,shift={(36,4)},line width =0.75pt] (1,-1)--(2,-2);
\draw [scale=0.12,shift={(36,4)},line width =0.75pt] (-1,-1)--(-2,-2);

\filldraw[draw,fill=red!100,scale=0.12,shift={(44,4)},line width =0.5pt] (-1,-1) rectangle (1,1);
\draw [scale=0.12,shift={(44,4)},line width =0.75pt] (1,1)--(2,2);
\draw [scale=0.12,shift={(44,4)},line width =0.75pt] (-1,1)--(-2,2);
\draw [scale=0.12,shift={(44,4)},line width =0.75pt] (1,-1)--(2,-2);
\draw [scale=0.12,shift={(44,4)},line width =0.75pt] (-1,-1)--(-2,-2);
\filldraw[draw,fill=red!100,scale=0.12,shift={(0,8)},line width =0.5pt] (-1,-1) rectangle (1,1);
\draw [scale=0.12,shift={(0,8)},line width =0.75pt] (1,1)--(2,2);
\draw [scale=0.12,shift={(0,8)},line width =0.75pt] (-1,1)--(-2,2);
\draw [scale=0.12,shift={(0,8)},line width =0.75pt] (1,-1)--(2,-2);
\draw [scale=0.12,shift={(0,8)},line width =0.75pt] (-1,-1)--(-2,-2);

\filldraw[draw,fill=red!100,scale=0.12,shift={(8,8)},line width =0.5pt] (-1,-1) rectangle (1,1);
\draw [scale=0.12,shift={(8,8)},line width =0.75pt] (1,1)--(2,2);
\draw [scale=0.12,shift={(8,8)},line width =0.75pt] (-1,1)--(-2,2);
\draw [scale=0.12,shift={(8,8)},line width =0.75pt] (1,-1)--(2,-2);
\draw [scale=0.12,shift={(8,8)},line width =0.75pt] (-1,-1)--(-2,-2);

\filldraw[draw,fill=red!100,scale=0.12,shift={(16,8)},line width =0.5pt] (-1,-1) rectangle (1,1);
\draw [scale=0.12,shift={(16,8)},line width =0.75pt] (1,1)--(2,2);
\draw [scale=0.12,shift={(16,8)},line width =0.75pt] (-1,1)--(-2,2);
\draw [scale=0.12,shift={(16,8)},line width =0.75pt] (1,-1)--(2,-2);
\draw [scale=0.12,shift={(16,8)},line width =0.75pt] (-1,-1)--(-2,-2);

\filldraw[draw,fill=red!100,scale=0.12,shift={(24,8)},line width =0.5pt] (-1,-1) rectangle (1,1);
\draw [scale=0.12,shift={(24,8)},line width =0.75pt] (1,1)--(2,2);
\draw [scale=0.12,shift={(24,8)},line width =0.75pt] (-1,1)--(-2,2);
\draw [scale=0.12,shift={(24,8)},line width =0.75pt] (1,-1)--(2,-2);
\draw [scale=0.12,shift={(24,8)},line width =0.75pt] (-1,-1)--(-2,-2);

\filldraw[draw,fill=red!100,scale=0.12,shift={(32,8)},line width =0.5pt] (-1,-1) rectangle (1,1);
\draw [scale=0.12,shift={(32,8)},line width =0.75pt] (1,1)--(2,2);
\draw [scale=0.12,shift={(32,8)},line width =0.75pt] (-1,1)--(-2,2);
\draw [scale=0.12,shift={(32,8)},line width =0.75pt] (1,-1)--(2,-2);
\draw [scale=0.12,shift={(32,8)},line width =0.75pt] (-1,-1)--(-2,-2);

\filldraw[draw,fill=red!100,scale=0.12,shift={(40,8)},line width =0.5pt] (-1,-1) rectangle (1,1);
\draw [scale=0.12,shift={(40,8)},line width =0.75pt] (1,1)--(2,2);
\draw [scale=0.12,shift={(40,8)},line width =0.75pt] (-1,1)--(-2,2);
\draw [scale=0.12,shift={(40,8)},line width =0.75pt] (1,-1)--(2,-2);
\draw [scale=0.12,shift={(40,8)},line width =0.75pt] (-1,-1)--(-2,-2);

\filldraw[draw,fill=red!100,scale=0.12,shift={(4,4+8)},line width =0.5pt] (-1,-1) rectangle (1,1);
\draw [scale=0.12,shift={(4,4+8)},line width =0.75pt] (1,1)--(2,2);
\draw [scale=0.12,shift={(4,4+8)},line width =0.75pt] (-1,1)--(-2,2);
\draw [scale=0.12,shift={(4,4+8)},line width =0.75pt] (1,-1)--(2,-2);
\draw [scale=0.12,shift={(4,4+8)},line width =0.75pt] (-1,-1)--(-2,-2);

\filldraw[draw,fill=red!100,scale=0.12,shift={(12,4+8)},line width =0.5pt] (-1,-1) rectangle (1,1);
\draw [scale=0.12,shift={(12,4+8)},line width =0.75pt] (1,1)--(2,2);
\draw [scale=0.12,shift={(12,4+8)},line width =0.75pt] (-1,1)--(-2,2);
\draw [scale=0.12,shift={(12,4+8)},line width =0.75pt] (1,-1)--(2,-2);
\draw [scale=0.12,shift={(12,4+8)},line width =0.75pt] (-1,-1)--(-2,-2);

\filldraw[draw,fill=red!100,scale=0.12,shift={(20,4+8)},line width =0.5pt] (-1,-1) rectangle (1,1);
\draw [scale=0.12,shift={(20,4+8)},line width =0.75pt] (1,1)--(2,2);
\draw [scale=0.12,shift={(20,4+8)},line width =0.75pt] (-1,1)--(-2,2);
\draw [scale=0.12,shift={(20,4+8)},line width =0.75pt] (1,-1)--(2,-2);
\draw [scale=0.12,shift={(20,4+8)},line width =0.75pt] (-1,-1)--(-2,-2);

\filldraw[draw,fill=red!100,scale=0.12,shift={(28,4+8)},line width =0.5pt] (-1,-1) rectangle (1,1);
\draw [scale=0.12,shift={(28,4+8)},line width =0.75pt] (1,1)--(2,2);
\draw [scale=0.12,shift={(28,4+8)},line width =0.75pt] (-1,1)--(-2,2);
\draw [scale=0.12,shift={(28,4+8)},line width =0.75pt] (1,-1)--(2,-2);
\draw [scale=0.12,shift={(28,4+8)},line width =0.75pt] (-1,-1)--(-2,-2);

\filldraw[draw,fill=red!100,scale=0.12,shift={(36,4+8)},line width =0.5pt] (-1,-1) rectangle (1,1);
\draw [scale=0.12,shift={(36,4+8)},line width =0.75pt] (1,1)--(2,2);
\draw [scale=0.12,shift={(36,4+8)},line width =0.75pt] (-1,1)--(-2,2);
\draw [scale=0.12,shift={(36,4+8)},line width =0.75pt] (1,-1)--(2,-2);
\draw [scale=0.12,shift={(36,4+8)},line width =0.75pt] (-1,-1)--(-2,-2);

\filldraw[draw,fill=red!100,scale=0.12,shift={(44,4+8)},line width =0.5pt] (-1,-1) rectangle (1,1);
\draw [scale=0.12,shift={(44,4+8)},line width =0.75pt] (1,1)--(2,2);
\draw [scale=0.12,shift={(44,4+8)},line width =0.75pt] (-1,1)--(-2,2);
\draw [scale=0.12,shift={(44,4+8)},line width =0.75pt] (1,-1)--(2,-2);
\draw [scale=0.12,shift={(44,4+8)},line width =0.75pt] (-1,-1)--(-2,-2);
\filldraw[draw,fill=red!100,scale=0.12,shift={(0,16)},line width =0.5pt] (-1,-1) rectangle (1,1);
\draw [scale=0.12,shift={(0,16)},line width =0.75pt] (1,1)--(2,2);
\draw [scale=0.12,shift={(0,16)},line width =0.75pt] (-1,1)--(-2,2);
\draw [scale=0.12,shift={(0,16)},line width =0.75pt] (1,-1)--(2,-2);
\draw [scale=0.12,shift={(0,16)},line width =0.75pt] (-1,-1)--(-2,-2);

\filldraw[draw,fill=red!100,scale=0.12,shift={(8,16)},line width =0.5pt] (-1,-1) rectangle (1,1);
\draw [scale=0.12,shift={(8,16)},line width =0.75pt] (1,1)--(2,2);
\draw [scale=0.12,shift={(8,16)},line width =0.75pt] (-1,1)--(-2,2);
\draw [scale=0.12,shift={(8,16)},line width =0.75pt] (1,-1)--(2,-2);
\draw [scale=0.12,shift={(8,16)},line width =0.75pt] (-1,-1)--(-2,-2);

\filldraw[draw,fill=red!100,scale=0.12,shift={(16,16)},line width =0.5pt] (-1,-1) rectangle (1,1);
\draw [scale=0.12,shift={(16,16)},line width =0.75pt] (1,1)--(2,2);
\draw [scale=0.12,shift={(16,16)},line width =0.75pt] (-1,1)--(-2,2);
\draw [scale=0.12,shift={(16,16)},line width =0.75pt] (1,-1)--(2,-2);
\draw [scale=0.12,shift={(16,16)},line width =0.75pt] (-1,-1)--(-2,-2);

\filldraw[draw,fill=red!100,scale=0.12,shift={(24,16)},line width =0.5pt] (-1,-1) rectangle (1,1);
\draw [scale=0.12,shift={(24,16)},line width =0.75pt] (1,1)--(2,2);
\draw [scale=0.12,shift={(24,16)},line width =0.75pt] (-1,1)--(-2,2);
\draw [scale=0.12,shift={(24,16)},line width =0.75pt] (1,-1)--(2,-2);
\draw [scale=0.12,shift={(24,16)},line width =0.75pt] (-1,-1)--(-2,-2);

\filldraw[draw,fill=red!100,scale=0.12,shift={(32,16)},line width =0.5pt] (-1,-1) rectangle (1,1);
\draw [scale=0.12,shift={(32,16)},line width =0.75pt] (1,1)--(2,2);
\draw [scale=0.12,shift={(32,16)},line width =0.75pt] (-1,1)--(-2,2);
\draw [scale=0.12,shift={(32,16)},line width =0.75pt] (1,-1)--(2,-2);
\draw [scale=0.12,shift={(32,16)},line width =0.75pt] (-1,-1)--(-2,-2);

\filldraw[draw,fill=red!100,scale=0.12,shift={(40,16)},line width =0.5pt] (-1,-1) rectangle (1,1);
\draw [scale=0.12,shift={(40,16)},line width =0.75pt] (1,1)--(2,2);
\draw [scale=0.12,shift={(40,16)},line width =0.75pt] (-1,1)--(-2,2);
\draw [scale=0.12,shift={(40,16)},line width =0.75pt] (1,-1)--(2,-2);
\draw [scale=0.12,shift={(40,16)},line width =0.75pt] (-1,-1)--(-2,-2);

\filldraw[draw,fill=red!100,scale=0.12,shift={(4,4+16)},line width =0.5pt] (-1,-1) rectangle (1,1);
\draw [scale=0.12,shift={(4,4+16)},line width =0.75pt] (1,1)--(2,2);
\draw [scale=0.12,shift={(4,4+16)},line width =0.75pt] (-1,1)--(-2,2);
\draw [scale=0.12,shift={(4,4+16)},line width =0.75pt] (1,-1)--(2,-2);
\draw [scale=0.12,shift={(4,4+16)},line width =0.75pt] (-1,-1)--(-2,-2);

\filldraw[draw,fill=red!100,scale=0.12,shift={(12,4+16)},line width =0.5pt] (-1,-1) rectangle (1,1);
\draw [scale=0.12,shift={(12,4+16)},line width =0.75pt] (1,1)--(2,2);
\draw [scale=0.12,shift={(12,4+16)},line width =0.75pt] (-1,1)--(-2,2);
\draw [scale=0.12,shift={(12,4+16)},line width =0.75pt] (1,-1)--(2,-2);
\draw [scale=0.12,shift={(12,4+16)},line width =0.75pt] (-1,-1)--(-2,-2);

\filldraw[draw,fill=red!100,scale=0.12,shift={(20,4+16)},line width =0.5pt] (-1,-1) rectangle (1,1);
\draw [scale=0.12,shift={(20,4+16)},line width =0.75pt] (1,1)--(2,2);
\draw [scale=0.12,shift={(20,4+16)},line width =0.75pt] (-1,1)--(-2,2);
\draw [scale=0.12,shift={(20,4+16)},line width =0.75pt] (1,-1)--(2,-2);
\draw [scale=0.12,shift={(20,4+16)},line width =0.75pt] (-1,-1)--(-2,-2);

\filldraw[draw,fill=red!100,scale=0.12,shift={(28,4+16)},line width =0.5pt] (-1,-1) rectangle (1,1);
\draw [scale=0.12,shift={(28,4+16)},line width =0.75pt] (1,1)--(2,2);
\draw [scale=0.12,shift={(28,4+16)},line width =0.75pt] (-1,1)--(-2,2);
\draw [scale=0.12,shift={(28,4+16)},line width =0.75pt] (1,-1)--(2,-2);
\draw [scale=0.12,shift={(28,4+16)},line width =0.75pt] (-1,-1)--(-2,-2);

\filldraw[draw,fill=red!100,scale=0.12,shift={(36,4+16)},line width =0.5pt] (-1,-1) rectangle (1,1);
\draw [scale=0.12,shift={(36,4+16)},line width =0.75pt] (1,1)--(2,2);
\draw [scale=0.12,shift={(36,4+16)},line width =0.75pt] (-1,1)--(-2,2);
\draw [scale=0.12,shift={(36,4+16)},line width =0.75pt] (1,-1)--(2,-2);
\draw [scale=0.12,shift={(36,4+16)},line width =0.75pt] (-1,-1)--(-2,-2);

\filldraw[draw,fill=red!100,scale=0.12,shift={(44,4+16)},line width =0.5pt] (-1,-1) rectangle (1,1);
\draw [scale=0.12,shift={(44,4+16)},line width =0.75pt] (1,1)--(2,2);
\draw [scale=0.12,shift={(44,4+16)},line width =0.75pt] (-1,1)--(-2,2);
\draw [scale=0.12,shift={(44,4+16)},line width =0.75pt] (1,-1)--(2,-2);
\draw [scale=0.12,shift={(44,4+16)},line width =0.75pt] (-1,-1)--(-2,-2);
\draw [fill,scale=0.12,shift={(18,22)},] (0,0) circle(.5);
\node at(1.9,2.6) {$\sigma_y^\beta$};
\filldraw[draw,fill=blue!100,scale=0.12,shift={(4,24)},line width =0.5pt] (-1,-1) rectangle (1,1);
\draw [scale=0.12,shift={(4,24)},line width =0.75pt] (1,1)--(2,2);
\draw [scale=0.12,shift={(4,24)},line width =0.75pt] (-1,1)--(-2,2);
\draw [scale=0.12,shift={(4,24)},line width =0.75pt] (1,-1)--(2,-2);
\draw [scale=0.12,shift={(4,24)},line width =0.75pt] (-1,-1)--(-2,-2);

\filldraw[draw,fill=blue!100,scale=0.12,shift={(12,24)},line width =0.5pt] (-1,-1) rectangle (1,1);
\draw [scale=0.12,shift={(12,24)},line width =0.75pt] (1,1)--(2,2);
\draw [scale=0.12,shift={(12,24)},line width =0.75pt] (-1,1)--(-2,2);
\draw [scale=0.12,shift={(12,24)},line width =0.75pt] (1,-1)--(2,-2);
\draw [scale=0.12,shift={(12,24)},line width =0.75pt] (-1,-1)--(-2,-2);

\filldraw[draw,fill=blue!100,scale=0.12,shift={(20,24)},line width =0.5pt] (-1,-1) rectangle (1,1);
\draw [scale=0.12,shift={(20,24)},line width =0.75pt] (1,1)--(2,2);
\draw [scale=0.12,shift={(20,24)},line width =0.75pt] (-1,1)--(-2,2);
\draw [scale=0.12,shift={(20,24)},line width =0.75pt] (1,-1)--(2,-2);
\draw [scale=0.12,shift={(20,24)},line width =0.75pt] (-1,-1)--(-2,-2);

\filldraw[draw,fill=blue!100,scale=0.12,shift={(28,24)},line width =0.5pt] (-1,-1) rectangle (1,1);
\draw [scale=0.12,shift={(28,24)},line width =0.75pt] (1,1)--(2,2);
\draw [scale=0.12,shift={(28,24)},line width =0.75pt] (-1,1)--(-2,2);
\draw [scale=0.12,shift={(28,24)},line width =0.75pt] (1,-1)--(2,-2);
\draw [scale=0.12,shift={(28,24)},line width =0.75pt] (-1,-1)--(-2,-2);

\filldraw[draw,fill=blue!100,scale=0.12,shift={(36,24)},line width =0.5pt] (-1,-1) rectangle (1,1);
\draw [scale=0.12,shift={(36,24)},line width =0.75pt] (1,1)--(2,2);
\draw [scale=0.12,shift={(36,24)},line width =0.75pt] (-1,1)--(-2,2);
\draw [scale=0.12,shift={(36,24)},line width =0.75pt] (1,-1)--(2,-2);
\draw [scale=0.12,shift={(36,24)},line width =0.75pt] (-1,-1)--(-2,-2);

\filldraw[draw,fill=blue!100,scale=0.12,shift={(44,24)},line width =0.5pt] (-1,-1) rectangle (1,1);
\draw [scale=0.12,shift={(44,24)},line width =0.75pt] (1,1)--(2,2);
\draw [scale=0.12,shift={(44,24)},line width =0.75pt] (-1,1)--(-2,2);
\draw [scale=0.12,shift={(44,24)},line width =0.75pt] (1,-1)--(2,-2);
\draw [scale=0.12,shift={(44,24)},line width =0.75pt] (-1,-1)--(-2,-2);

\filldraw[draw,fill=blue!100,scale=0.12,shift={(0,28)},line width =0.5pt] (-1,-1) rectangle (1,1);
\draw [scale=0.12,shift={(0,28)},line width =0.75pt] (1,1)--(2,2);
\draw [scale=0.12,shift={(0,28)},line width =0.75pt] (-1,1)--(-2,2);
\draw [scale=0.12,shift={(0,28)},line width =0.75pt] (1,-1)--(2,-2);
\draw [scale=0.12,shift={(0,28)},line width =0.75pt] (-1,-1)--(-2,-2);

\filldraw[draw,fill=blue!100,scale=0.12,shift={(8,28)},line width =0.5pt] (-1,-1) rectangle (1,1);
\draw [scale=0.12,shift={(8,28)},line width =0.75pt] (1,1)--(2,2);
\draw [scale=0.12,shift={(8,28)},line width =0.75pt] (-1,1)--(-2,2);
\draw [scale=0.12,shift={(8,28)},line width =0.75pt] (1,-1)--(2,-2);
\draw [scale=0.12,shift={(8,28)},line width =0.75pt] (-1,-1)--(-2,-2);

\filldraw[draw,fill=blue!100,scale=0.12,shift={(16,28)},line width =0.5pt] (-1,-1) rectangle (1,1);
\draw [scale=0.12,shift={(16,28)},line width =0.75pt] (1,1)--(2,2);
\draw [scale=0.12,shift={(16,28)},line width =0.75pt] (-1,1)--(-2,2);
\draw [scale=0.12,shift={(16,28)},line width =0.75pt] (1,-1)--(2,-2);
\draw [scale=0.12,shift={(16,28)},line width =0.75pt] (-1,-1)--(-2,-2);

\filldraw[draw,fill=blue!100,scale=0.12,shift={(24,28)},line width =0.5pt] (-1,-1) rectangle (1,1);
\draw [scale=0.12,shift={(24,28)},line width =0.75pt] (1,1)--(2,2);
\draw [scale=0.12,shift={(24,28)},line width =0.75pt] (-1,1)--(-2,2);
\draw [scale=0.12,shift={(24,28)},line width =0.75pt] (1,-1)--(2,-2);
\draw [scale=0.12,shift={(24,28)},line width =0.75pt] (-1,-1)--(-2,-2);

\filldraw[draw,fill=blue!100,scale=0.12,shift={(32,28)},line width =0.5pt] (-1,-1) rectangle (1,1);
\draw [scale=0.12,shift={(32,28)},line width =0.75pt] (1,1)--(2,2);
\draw [scale=0.12,shift={(32,28)},line width =0.75pt] (-1,1)--(-2,2);
\draw [scale=0.12,shift={(32,28)},line width =0.75pt] (1,-1)--(2,-2);
\draw [scale=0.12,shift={(32,28)},line width =0.75pt] (-1,-1)--(-2,-2);

\filldraw[draw,fill=blue!100,scale=0.12,shift={(40,28)},line width =0.5pt] (-1,-1) rectangle (1,1);
\draw [scale=0.12,shift={(40,28)},line width =0.75pt] (1,1)--(2,2);
\draw [scale=0.12,shift={(40,28)},line width =0.75pt] (-1,1)--(-2,2);
\draw [scale=0.12,shift={(40,28)},line width =0.75pt] (1,-1)--(2,-2);
\draw [scale=0.12,shift={(40,28)},line width =0.75pt] (-1,-1)--(-2,-2);
\filldraw[draw,fill=blue!100,scale=0.12,shift={(4,24+8)},line width =0.5pt] (-1,-1) rectangle (1,1);
\draw [scale=0.12,shift={(4,24+8)},line width =0.75pt] (1,1)--(2,2);
\draw [scale=0.12,shift={(4,24+8)},line width =0.75pt] (-1,1)--(-2,2);
\draw [scale=0.12,shift={(4,24+8)},line width =0.75pt] (1,-1)--(2,-2);
\draw [scale=0.12,shift={(4,24+8)},line width =0.75pt] (-1,-1)--(-2,-2);

\filldraw[draw,fill=blue!100,scale=0.12,shift={(12,24+8)},line width =0.5pt] (-1,-1) rectangle (1,1);
\draw [scale=0.12,shift={(12,24+8)},line width =0.75pt] (1,1)--(2,2);
\draw [scale=0.12,shift={(12,24+8)},line width =0.75pt] (-1,1)--(-2,2);
\draw [scale=0.12,shift={(12,24+8)},line width =0.75pt] (1,-1)--(2,-2);
\draw [scale=0.12,shift={(12,24+8)},line width =0.75pt] (-1,-1)--(-2,-2);

\filldraw[draw,fill=blue!100,scale=0.12,shift={(20,24+8)},line width =0.5pt] (-1,-1) rectangle (1,1);
\draw [scale=0.12,shift={(20,24+8)},line width =0.75pt] (1,1)--(2,2);
\draw [scale=0.12,shift={(20,24+8)},line width =0.75pt] (-1,1)--(-2,2);
\draw [scale=0.12,shift={(20,24+8)},line width =0.75pt] (1,-1)--(2,-2);
\draw [scale=0.12,shift={(20,24+8)},line width =0.75pt] (-1,-1)--(-2,-2);

\filldraw[draw,fill=blue!100,scale=0.12,shift={(28,24+8)},line width =0.5pt] (-1,-1) rectangle (1,1);
\draw [scale=0.12,shift={(28,24+8)},line width =0.75pt] (1,1)--(2,2);
\draw [scale=0.12,shift={(28,24+8)},line width =0.75pt] (-1,1)--(-2,2);
\draw [scale=0.12,shift={(28,24+8)},line width =0.75pt] (1,-1)--(2,-2);
\draw [scale=0.12,shift={(28,24+8)},line width =0.75pt] (-1,-1)--(-2,-2);

\filldraw[draw,fill=blue!100,scale=0.12,shift={(36,24+8)},line width =0.5pt] (-1,-1) rectangle (1,1);
\draw [scale=0.12,shift={(36,24+8)},line width =0.75pt] (1,1)--(2,2);
\draw [scale=0.12,shift={(36,24+8)},line width =0.75pt] (-1,1)--(-2,2);
\draw [scale=0.12,shift={(36,24+8)},line width =0.75pt] (1,-1)--(2,-2);
\draw [scale=0.12,shift={(36,24+8)},line width =0.75pt] (-1,-1)--(-2,-2);

\filldraw[draw,fill=blue!100,scale=0.12,shift={(44,24+8)},line width =0.5pt] (-1,-1) rectangle (1,1);
\draw [scale=0.12,shift={(44,24+8)},line width =0.75pt] (1,1)--(2,2);
\draw [scale=0.12,shift={(44,24+8)},line width =0.75pt] (-1,1)--(-2,2);
\draw [scale=0.12,shift={(44,24+8)},line width =0.75pt] (1,-1)--(2,-2);
\draw [scale=0.12,shift={(44,24+8)},line width =0.75pt] (-1,-1)--(-2,-2);

\filldraw[draw,fill=blue!100,scale=0.12,shift={(0,28+8)},line width =0.5pt] (-1,-1) rectangle (1,1);
\draw [scale=0.12,shift={(0,28+8)},line width =0.75pt] (1,1)--(2,2);
\draw [scale=0.12,shift={(0,28+8)},line width =0.75pt] (-1,1)--(-2,2);
\draw [scale=0.12,shift={(0,28+8)},line width =0.75pt] (1,-1)--(2,-2);
\draw [scale=0.12,shift={(0,28+8)},line width =0.75pt] (-1,-1)--(-2,-2);

\filldraw[draw,fill=blue!100,scale=0.12,shift={(8,28+8)},line width =0.5pt] (-1,-1) rectangle (1,1);
\draw [scale=0.12,shift={(8,28+8)},line width =0.75pt] (1,1)--(2,2);
\draw [scale=0.12,shift={(8,28+8)},line width =0.75pt] (-1,1)--(-2,2);
\draw [scale=0.12,shift={(8,28+8)},line width =0.75pt] (1,-1)--(2,-2);
\draw [scale=0.12,shift={(8,28+8)},line width =0.75pt] (-1,-1)--(-2,-2);

\filldraw[draw,fill=blue!100,scale=0.12,shift={(16,28+8)},line width =0.5pt] (-1,-1) rectangle (1,1);
\draw [scale=0.12,shift={(16,28+8)},line width =0.75pt] (1,1)--(2,2);
\draw [scale=0.12,shift={(16,28+8)},line width =0.75pt] (-1,1)--(-2,2);
\draw [scale=0.12,shift={(16,28+8)},line width =0.75pt] (1,-1)--(2,-2);
\draw [scale=0.12,shift={(16,28+8)},line width =0.75pt] (-1,-1)--(-2,-2);

\filldraw[draw,fill=blue!100,scale=0.12,shift={(24,28+8)},line width =0.5pt] (-1,-1) rectangle (1,1);
\draw [scale=0.12,shift={(24,28+8)},line width =0.75pt] (1,1)--(2,2);
\draw [scale=0.12,shift={(24,28+8)},line width =0.75pt] (-1,1)--(-2,2);
\draw [scale=0.12,shift={(24,28+8)},line width =0.75pt] (1,-1)--(2,-2);
\draw [scale=0.12,shift={(24,28+8)},line width =0.75pt] (-1,-1)--(-2,-2);

\filldraw[draw,fill=blue!100,scale=0.12,shift={(32,28+8)},line width =0.5pt] (-1,-1) rectangle (1,1);
\draw [scale=0.12,shift={(32,28+8)},line width =0.75pt] (1,1)--(2,2);
\draw [scale=0.12,shift={(32,28+8)},line width =0.75pt] (-1,1)--(-2,2);
\draw [scale=0.12,shift={(32,28+8)},line width =0.75pt] (1,-1)--(2,-2);
\draw [scale=0.12,shift={(32,28+8)},line width =0.75pt] (-1,-1)--(-2,-2);

\filldraw[draw,fill=blue!100,scale=0.12,shift={(40,28+8)},line width =0.5pt] (-1,-1) rectangle (1,1);
\draw [scale=0.12,shift={(40,28+8)},line width =0.75pt] (1,1)--(2,2);
\draw [scale=0.12,shift={(40,28+8)},line width =0.75pt] (-1,1)--(-2,2);
\draw [scale=0.12,shift={(40,28+8)},line width =0.75pt] (1,-1)--(2,-2);
\draw [scale=0.12,shift={(40,28+8)},line width =0.75pt] (-1,-1)--(-2,-2);
\filldraw[draw,fill=blue!100,scale=0.12,shift={(4,24+16)},line width =0.5pt] (-1,-1) rectangle (1,1);
\draw [scale=0.12,shift={(4,24+16)},line width =0.75pt] (1,1)--(2,2);
\draw [scale=0.12,shift={(4,24+16)},line width =0.75pt] (-1,1)--(-2,2);
\draw [scale=0.12,shift={(4,24+16)},line width =0.75pt] (1,-1)--(2,-2);
\draw [scale=0.12,shift={(4,24+16)},line width =0.75pt] (-1,-1)--(-2,-2);

\filldraw[draw,fill=blue!100,scale=0.12,shift={(12,24+16)},line width =0.5pt] (-1,-1) rectangle (1,1);
\draw [scale=0.12,shift={(12,24+16)},line width =0.75pt] (1,1)--(2,2);
\draw [scale=0.12,shift={(12,24+16)},line width =0.75pt] (-1,1)--(-2,2);
\draw [scale=0.12,shift={(12,24+16)},line width =0.75pt] (1,-1)--(2,-2);
\draw [scale=0.12,shift={(12,24+16)},line width =0.75pt] (-1,-1)--(-2,-2);

\filldraw[draw,fill=blue!100,scale=0.12,shift={(20,24+16)},line width =0.5pt] (-1,-1) rectangle (1,1);
\draw [scale=0.12,shift={(20,24+16)},line width =0.75pt] (1,1)--(2,2);
\draw [scale=0.12,shift={(20,24+16)},line width =0.75pt] (-1,1)--(-2,2);
\draw [scale=0.12,shift={(20,24+16)},line width =0.75pt] (1,-1)--(2,-2);
\draw [scale=0.12,shift={(20,24+16)},line width =0.75pt] (-1,-1)--(-2,-2);

\filldraw[draw,fill=blue!100,scale=0.12,shift={(28,24+16)},line width =0.5pt] (-1,-1) rectangle (1,1);
\draw [scale=0.12,shift={(28,24+16)},line width =0.75pt] (1,1)--(2,2);
\draw [scale=0.12,shift={(28,24+16)},line width =0.75pt] (-1,1)--(-2,2);
\draw [scale=0.12,shift={(28,24+16)},line width =0.75pt] (1,-1)--(2,-2);
\draw [scale=0.12,shift={(28,24+16)},line width =0.75pt] (-1,-1)--(-2,-2);

\filldraw[draw,fill=blue!100,scale=0.12,shift={(36,24+16)},line width =0.5pt] (-1,-1) rectangle (1,1);
\draw [scale=0.12,shift={(36,24+16)},line width =0.75pt] (1,1)--(2,2);
\draw [scale=0.12,shift={(36,24+16)},line width =0.75pt] (-1,1)--(-2,2);
\draw [scale=0.12,shift={(36,24+16)},line width =0.75pt] (1,-1)--(2,-2);
\draw [scale=0.12,shift={(36,24+16)},line width =0.75pt] (-1,-1)--(-2,-2);

\filldraw[draw,fill=blue!100,scale=0.12,shift={(44,24+16)},line width =0.5pt] (-1,-1) rectangle (1,1);
\draw [scale=0.12,shift={(44,24+16)},line width =0.75pt] (1,1)--(2,2);
\draw [scale=0.12,shift={(44,24+16)},line width =0.75pt] (-1,1)--(-2,2);
\draw [scale=0.12,shift={(44,24+16)},line width =0.75pt] (1,-1)--(2,-2);
\draw [scale=0.12,shift={(44,24+16)},line width =0.75pt] (-1,-1)--(-2,-2);

\filldraw[draw,fill=blue!100,scale=0.12,shift={(0,28+16)},line width =0.5pt] (-1,-1) rectangle (1,1);
\draw [scale=0.12,shift={(0,28+16)},line width =0.75pt] (1,1)--(2,2);
\draw [scale=0.12,shift={(0,28+16)},line width =0.75pt] (-1,1)--(-2,2);
\draw [scale=0.12,shift={(0,28+16)},line width =0.75pt] (1,-1)--(2,-2);
\draw [scale=0.12,shift={(0,28+16)},line width =0.75pt] (-1,-1)--(-2,-2);

\filldraw[draw,fill=blue!100,scale=0.12,shift={(8,28+16)},line width =0.5pt] (-1,-1) rectangle (1,1);
\draw [scale=0.12,shift={(8,28+16)},line width =0.75pt] (1,1)--(2,2);
\draw [scale=0.12,shift={(8,28+16)},line width =0.75pt] (-1,1)--(-2,2);
\draw [scale=0.12,shift={(8,28+16)},line width =0.75pt] (1,-1)--(2,-2);
\draw [scale=0.12,shift={(8,28+16)},line width =0.75pt] (-1,-1)--(-2,-2);

\filldraw[draw,fill=blue!100,scale=0.12,shift={(16,28+16)},line width =0.5pt] (-1,-1) rectangle (1,1);
\draw [scale=0.12,shift={(16,28+16)},line width =0.75pt] (1,1)--(2,2);
\draw [scale=0.12,shift={(16,28+16)},line width =0.75pt] (-1,1)--(-2,2);
\draw [scale=0.12,shift={(16,28+16)},line width =0.75pt] (1,-1)--(2,-2);
\draw [scale=0.12,shift={(16,28+16)},line width =0.75pt] (-1,-1)--(-2,-2);

\filldraw[draw,fill=blue!100,scale=0.12,shift={(24,28+16)},line width =0.5pt] (-1,-1) rectangle (1,1);
\draw [scale=0.12,shift={(24,28+16)},line width =0.75pt] (1,1)--(2,2);
\draw [scale=0.12,shift={(24,28+16)},line width =0.75pt] (-1,1)--(-2,2);
\draw [scale=0.12,shift={(24,28+16)},line width =0.75pt] (1,-1)--(2,-2);
\draw [scale=0.12,shift={(24,28+16)},line width =0.75pt] (-1,-1)--(-2,-2);

\filldraw[draw,fill=blue!100,scale=0.12,shift={(32,28+16)},line width =0.5pt] (-1,-1) rectangle (1,1);
\draw [scale=0.12,shift={(32,28+16)},line width =0.75pt] (1,1)--(2,2);
\draw [scale=0.12,shift={(32,28+16)},line width =0.75pt] (-1,1)--(-2,2);
\draw [scale=0.12,shift={(32,28+16)},line width =0.75pt] (1,-1)--(2,-2);
\draw [scale=0.12,shift={(32,28+16)},line width =0.75pt] (-1,-1)--(-2,-2);

\filldraw[draw,fill=blue!100,scale=0.12,shift={(40,28+16)},line width =0.5pt] (-1,-1) rectangle (1,1);
\draw [scale=0.12,shift={(40,28+16)},line width =0.75pt] (1,1)--(2,2);
\draw [scale=0.12,shift={(40,28+16)},line width =0.75pt] (-1,1)--(-2,2);
\draw [scale=0.12,shift={(40,28+16)},line width =0.75pt] (1,-1)--(2,-2);
\draw [scale=0.12,shift={(40,28+16)},line width =0.75pt] (-1,-1)--(-2,-2);
\draw [fill,scale=0.12,shift={(38,46)},] (0,0) circle(.5);
\node at(4.3,5.3) {$\sigma_x^\alpha$};
\end{tikzpicture}
\end{equation}
where the boundary condition of space is periodic and the trace of the equation~\eqref{DCF} is realized by the periodic boundary condition in time direction.

One can exploit the properties of dual-unitary for the exact computation of dynamic correlation functions.
As dual-unitary circuits are unitary in both time and space directions, the contraction is also implemented for two ends of the horizon direction of the circuit under a periodic boundary condition. Moreover, two ends for the vertical direction are also contracted, as the infinite temperature state is proportional to an identity matrix.
Those properties cause the vast majority of blocks to be eliminated in the calculation.
If two single local operators are not both located at the edge of the light-cone, the result of contraction will always have terms that trace the single local Pauli operator, which results in zero.
Thus, the dynamical correlation functions of local operators under an infinite temperature state are nonzero only on the edges of the light-cone as given by Ref.~\cite{BrunoBertini2019}.

Remarkably, the light-cone correlation functions of local operators under an infinite temperature state can be exactly calculated by the transfer matrix method. The transfer matrix $\mathcal{M}_\pm(a)$ describes a transfer of single-site operator along the two opposite directions of the light-cone. $\mathcal{M}_\pm(a)$ are linear mapping superoperators, which can be represented graphically as,
\begin{equation}
\begin{tikzpicture}
\node at(-1,0) {$\mathcal{M}_+(a)=\frac{1}{2}\text{tr}_1[U^\dag(a \otimes \mathbbm{1}_2)U]=\frac{1}{2}$};
\filldraw[draw,fill=blue!100,scale=0.20,shift={(12,3)},line width =0.75pt] (-1,-1) rectangle (1,1);
\draw [scale=0.20,shift={(12,3)},line width =1pt] (-1,1)--(-3,1);
\draw [scale=0.20,shift={(12,3)},line width =1pt] (1,1)--(2,2);

\draw [scale=0.20,shift={(12,3)},line width =1pt] (1,-1)--(1,-5);
\draw [scale=0.20,shift={(12,3)},line width =1pt] (-1,-1)--(-1,-5);
\draw [fill,scale=0.20,shift={(12,2)}] (-1,-2) circle(.4);
\draw [scale=0.20,shift={(12,3)},line width =1pt] (-3,1)--(-4,0)--(-4,-6)--(-3,-7);
\node at(1.9,0) {$a$};

\filldraw[draw,fill=red!100,scale=0.20,shift={(12,-3)},line width =0.75pt] (-1,-1) rectangle (1,1);
\draw [scale=0.20,shift={(12,-3)},line width =1pt] (-1,-1)--(-3,-1);
\draw [scale=0.20,shift={(12,-3)},line width =1pt] (1,-1)--(2,-2);
\node at(3.0,0) {,};
\end{tikzpicture}
\end{equation}
and,
\begin{equation}
\begin{tikzpicture}
\node at(0.07,0) {$\mathcal{M}_-(a)=\frac{1}{2}\text{tr}_2[U^\dag(\mathbbm{1}_2 \otimes a)U]=\frac{1}{2}$};
\filldraw[draw,fill=blue!100,scale=0.20,shift={(16,3)},line width =0.75pt] (-1,-1) rectangle (1,1);
\draw [scale=0.20,shift={(16,3)},line width =1pt] (-1,1)--(-2,2);
\draw [scale=0.20,shift={(16,3)},line width =1pt] (1,1)--(3,1);

\draw [scale=0.20,shift={(16,3)},line width =1pt] (1,-1)--(1,-5);
\draw [scale=0.20,shift={(16,3)},line width =1pt] (-1,-1)--(-1,-5);
\draw [fill,scale=0.20,shift={(16,3)}] (1,-3) circle(.4);
\draw [scale=0.20,shift={(16,3)},line width =1pt] (3,1)--(4,0)--(4,-6)--(3,-7);
\node at(3.7,0) {$a$};

\filldraw[draw,fill=red!100,scale=0.20,shift={(16,-3)},line width =0.75pt] (-1,-1) rectangle (1,1);
\draw [scale=0.20,shift={(16,-3)},line width =1pt] (-1,-1)--(-2,-2);
\draw [scale=0.20,shift={(16,-3)},line width =1pt] (1,-1)--(3,-1);
\node at(4.2,0) {.};
\end{tikzpicture}
\end{equation}
Typically, one can choose the operator $a$ as a Pauli matrix. For qubits $\mathcal{M}_\pm(a)$ is a $4\times4$ matrix. When implementing $\mathcal{M}_\pm(a)$, a Pauli matrix at one site will be mapped into a mixture of all Pauli matrices on the neighbor site. In this regard, the spreading of information along the edge of the light-cone can be well captured by the property of the transfer matrix  $\mathcal{M}_\pm(a)$.

For two separated local operators at the light-cone, the dynamical correlation function can be calculated using the transfer matrix as following,
\begin{equation}\label{LCCLFs}
	\begin{split}
		C^{\alpha,\beta}_\pm(r=\pm t,t)=&\frac{1}{2}\text{tr}[\sigma^\alpha\mathcal{M}^{2t}_\pm(\sigma^\beta)],\\
	\end{split}
\end{equation}
where $r$ is the separation of the two local operators in the space. The diagrammatical representation of equation \eqref{LCCLFs} is shown as,
\begin{equation}
\begin{tikzpicture}
\node at(-2.4,0) {$C^{\alpha,\beta}_+(t,t)=\frac{1}{2^{2t+1}}$};
\filldraw[draw,fill=blue!100,scale=0.12,shift={(15,2)},line width =0.5pt] (-1,-1) rectangle (1,1);
\draw [scale=0.12,shift={(15,2)},line width =0.75pt] (-1,1)--(-4,1);
\draw [scale=0.12,shift={(15,2)},line width =0.75pt] (1,1)--(2,2);

\draw [scale=0.12,shift={(15,2)},line width =0.75pt] (1,-1)--(1,-3);
\draw [scale=0.12,shift={(15,2)},line width =0.75pt] (-1,-1)--(-1,-3);
\draw [fill,scale=0.12,shift={(15,2)}] (-1,-2) circle(.4);
\draw [scale=0.12,shift={(15,2)},line width =0.75pt] (-4,1)--(-5,0)--(-5,-4)--(-4,-5);
\node at(1.45,0.1) {$\sigma^\beta$};

\filldraw[draw,fill=red!100,scale=0.12,shift={(15,-2)},line width =0.5pt] (-1,-1) rectangle (1,1);
\draw [scale=0.12,shift={(15,-2)},line width =0.75pt] (-1,-1)--(-4,-1);
\draw [scale=0.12,shift={(15,-2)},line width =0.75pt] (1,-1)--(2,-2);

\filldraw[draw,fill=blue!100,scale=0.12,shift={(15+4,2+4)},line width =0.5pt] (-1,-1) rectangle (1,1);
\draw [scale=0.12,shift={(15+4,2+4)},line width =0.75pt] (-1,-1)--(-2.1,-2.1);
\draw [scale=0.12,shift={(15+4,2+4)},line width =0.75pt] (-1,1)--(-2-8,1);
\draw [scale=0.12,shift={(15+4,2+4)},line width =0.75pt] (1,1)--(2.1,2.1);

\draw [scale=0.12,shift={(15+4,2+4)},line width =0.75pt] (-2-8,1)--(-4-8,-1)--(-8-4,-8-3)--(-2-8,-5-8);
\draw [scale=0.12,shift={(15+4,2+4)},line width =0.75pt] (1,-1)--(1,-1-2-8);

\filldraw[draw,fill=red!100,scale=0.12,shift={(15+4,-2-4)},line width =0.5pt] (-1,-1) rectangle (1,1);
\draw [scale=0.12,shift={(15+4,-2-4)},line width =0.75pt] (-1,-1)--(-2-8,-1);
\draw [scale=0.12,shift={(15+4,-2-4)},line width =0.75pt] (-1,1)--(-2.1,2.1);
\draw [scale=0.12,shift={(15+4,-2-4)},line width =0.75pt] (1,-1)--(2.1,-2.1);

\filldraw[draw,fill=blue!100,scale=0.12,shift={(15+4*2,2+4*2)},line width =0.5pt] (-1,-1) rectangle (1,1);
\draw [scale=0.12,shift={(15+4*2,2+4*2)},line width =0.75pt] (-1,-1)--(-2.1,-2.1);
\draw [scale=0.12,shift={(15+4*2,2+4*2)},line width =0.75pt] (-1,1)--(-2-8*2,1);
\draw [scale=0.12,shift={(15+4*2,2+4*2)},line width =0.75pt] (1,1)--(2.1,2.1);

\draw [scale=0.12,shift={(15+4*2,2+4*2)},line width =0.75pt] (-2-8*2,1)--(-4-8*2,-1)--(-4-8*2,-3-8*2)--(-2-8*2,-5-8*2);
\draw [scale=0.12,shift={(15+4*2,2+4*2)},line width =0.75pt] (1,-1)--(1,-1-2-8*2);

\filldraw[draw,fill=red!100,scale=0.12,shift={(15+4*2,-2-4*2)},line width =0.5pt] (-1,-1) rectangle (1,1);
\draw [scale=0.12,shift={(15+4*2,-2-4*2)},line width =0.75pt] (-1,-1)--(-2-8*2,-1);
\draw [scale=0.12,shift={(15+4*2,-2-4*2)},line width =0.75pt] (-1,1)--(-2.1,2.1);
\draw [scale=0.12,shift={(15+4*2,-2-4*2)},line width =0.75pt] (1,-1)--(2.1,-2.1);

\filldraw[draw,fill=blue!100,scale=0.12,shift={(15+4*3,2+4*3)},line width =0.5pt] (-1,-1) rectangle (1,1);
\draw [scale=0.12,shift={(15+4*3,2+4*3)},line width =0.75pt] (-1,-1)--(-2.1,-2.1);
\draw [scale=0.12,shift={(15+4*3,2+4*3)},line width =0.75pt] (-1,1)--(-2-8*3,1);
\draw [scale=0.12,shift={(15+4*3,2+4*3)},line width =0.75pt] (1,1)--(2.1,2.1);

\draw [scale=0.12,shift={(15+4*3,2+4*3)},line width =0.75pt] (-2-8*3,1)--(-4-8*3,-1)--(-4-8*3,-3-8*3)--(-2-8*3,-5-8*3);
\draw [scale=0.12,shift={(15+4*3,2+4*3)},line width =0.75pt] (1,-1)--(1,-1-2-8*3);

\filldraw[draw,fill=red!100,scale=0.12,shift={(15+4*3,-2-4*3)},line width =0.5pt] (-1,-1) rectangle (1,1);
\draw [scale=0.12,shift={(15+4*3,-2-4*3)},line width =0.75pt] (-1,-1)--(-2-8*3,-1);
\draw [scale=0.12,shift={(15+4*3,-2-4*3)},line width =0.75pt] (-1,1)--(-2.1,2.1);
\draw [scale=0.12,shift={(15+4*3,-2-4*3)},line width =0.75pt] (1,-1)--(2.1,-2.1);

\filldraw[draw,fill=blue!100,scale=0.12,shift={(15+4*4,2+4*4)},line width =0.5pt] (-1,-1) rectangle (1,1);
\draw [scale=0.12,shift={(15+4*4,2+4*4)},line width =0.75pt] (-1,-1)--(-2.1,-2.1);
\draw [scale=0.12,shift={(15+4*4,2+4*4)},line width =0.75pt] (-1,1)--(-2-8*4,1);
\draw [scale=0.12,shift={(15+4*4,2+4*4)},line width =0.75pt] (1,1)--(2.1,2.1);

\draw [scale=0.12,shift={(15+4*4,2+4*4)},line width =0.75pt] (-2-8*4,1)--(-4-8*4,-1)--(-4-8*4,-3-8*4)--(-2-8*4,-5-8*4);
\draw [scale=0.12,shift={(15+4*4,2+4*4)},line width =0.75pt] (1,-1)--(1,-1-2-8*4);

\filldraw[draw,fill=red!100,scale=0.12,shift={(15+4*4,-2-4*4)},line width =0.5pt] (-1,-1) rectangle (1,1);
\draw [scale=0.12,shift={(15+4*4,-2-4*4)},line width =0.75pt] (-1,-1)--(-2-8*4,-1);
\draw [scale=0.12,shift={(15+4*4,-2-4*4)},line width =0.75pt] (-1,1)--(-2.1,2.1);
\draw [scale=0.12,shift={(15+4*4,-2-4*4)},line width =0.75pt] (1,-1)--(2.1,-2.1);

\filldraw[draw,fill=blue!100,scale=0.12,shift={(15+4*5,2+4*5)},line width =0.5pt] (-1,-1) rectangle (1,1);
\draw [scale=0.12,shift={(15+4*5,2+4*5)},line width =0.75pt] (-1,-1)--(-2.1,-2.1);
\draw [scale=0.12,shift={(15+4*5,2+4*5)},line width =0.75pt] (-1,1)--(-2-8*5,1);
\draw [scale=0.12,shift={(15+4*5,2+4*5)},line width =0.75pt] (1,1)--(2.1+1,2.1+1);

\draw [scale=0.12,shift={(15+4*5,2+4*5)},line width =0.75pt] (-2-8*5,1)--(-4-8*5,-1)--(-4-8*5,-3-8*5)--(-2-8*5,-5-8*5);
\draw [scale=0.12,shift={(15+4*5,2+4*5)},line width =0.75pt] (1,-1)--(1,-1-2-8*5);

\filldraw[draw,fill=red!100,scale=0.12,shift={(15+4*5,-2-4*5)},line width =0.5pt] (-1,-1) rectangle (1,1);
\draw [scale=0.12,shift={(15+4*5,-2-4*5)},line width =0.75pt] (-1,-1)--(-2-8*5,-1);
\draw [scale=0.12,shift={(15+4*5,-2-4*5)},line width =0.75pt] (-1,1)--(-2.1,2.1);
\draw [scale=0.12,shift={(15+4*5,-2-4*5)},line width =0.75pt] (1,-1)--(2.1+1,-2.1-1)--(2.1+1,47);

\draw [fill,scale=0.12,shift={(0,0)}] (38,25) circle(.4);
\node at(36*0.12,25.5*0.12) {$\sigma^\alpha$};
\node at(39*0.12,0*0.12) {.};
\end{tikzpicture}
\end{equation}
The graph of $C^{\alpha,\beta}_-(-t,t)$ has a similar structure to $C^{\alpha,\beta}_+(t,t)$.
Moreover, the properties of $C^{\alpha,\beta}_-(-t,t)$ are also similar to $C^{\alpha,\beta}_+(t,t)$ without qualitative differences. Without loss of generality, our investigation only shows the results of $C^{\alpha,\beta}_+(t,t)$.
Ref.~\cite{BrunoBertini2019} shows a classification of dual-unitary circuits of qubits by analyzing the eigenvalues of the transfer matrix as follows:

(i) Noninteracting behavior: all eigenvalues $\lambda_i=1$, which indicate all dynamical correlations show constant with time.

(ii) Nonergodic (and generically interacting and non-integrable) behavior: part of the eigenvalue $\lambda_i=1$, which implies some dynamical correlations remain constant with time.

(iii) Ergodic but nonmixing behavior: all nontrivial eigenvalues $\lambda_i\neq1$, but there is at least one eigenvalue with the unit modulus $|\lambda_i|=1$.
In this case, all time-averaged dynamical correlations will vanish after a long-time evolution.

(iv) Ergodic and mixing behavior: all nontrivial eigenvalues are within the unit disk, namely $|\lambda|<1$. In this case, all dynamic correlations will vanish after a long time without time-averaged.

Here, $\lambda_i$ is the eigenvalue of the transfer matrix $\mathcal{M}$ with $i=0,1,2,3$.

The classification only considers eigenvalues of the transfer matrix $\mathcal{M}_\pm(a)$. However, the transfer matrix in general is not Hermitian and may contain Jordan blocks. In other words, the transfer matrix may have EPs. Consequently, the dynamical correlation function should have a polynomial-enhanced behavior. Although this has been noticed in Ref.~\cite{BrunoBertini2019}, the construction of a dual-unitary quantum circuit where the transfer matrix has EPs has not been explicitly given, leading to a lack of a full classification of dual-unitary quantum circuits. In additional, it is noted that consequences of non-diagonalizable transfer matrix also show up for quantum transport properties~\cite{MadhumitaSaha2023a,MadhumitaSaha2023b,MadhumitaSaha2023c}, although not in the context of dual-unitary quantum circuit.

\section{Transfer matrix with EPs}
\label{s3}

In this section, we give some explicit constructions of dual-unitary quantum circuits where the transfer matrices can have EPs. Then, we investigate the behavior of dynamic correlation functions.

The goal can be set as to find a dual-unitary two-qubit gate $U$ whose corresponding transfer matrix has EPs. Our strategy is to parameterize both the transfer matrix and the dual-unitary $U$, and then derive a series of equations that can make constraints for the solutions. We find that there is a one-parameter family of dual-unitary quantum circuits where the transfer matrix has EPs.

Although the parameterization of general dual-unitary gates are still not available, the dual-unitary two-qubit gate can be completely parameterized as~\cite{BrunoBertini2019},
\begin{equation}\label{}
U=e^{i\theta}(u_+\otimes u_-)V[J](v_+\otimes v_-),
\end{equation}
where $\theta \in \mathbbm{R}$, $u_\pm, v_\pm \in \text{SU(2)}$ is a single-qubit gate, and,
\begin{equation}
V[J]=e^{-i(\frac{\pi}{4}\sigma^x\otimes\sigma^x+\frac{\pi}{4}\sigma^y\otimes\sigma^y+J\sigma^z\otimes\sigma^z)}.
\end{equation}
For this case of the dual-unitary circuit, the transfer matrix is $4\times4$ matrix. Here, we chose the mapping,
\begin{equation}
\sigma_0\rightarrow{\begin{pmatrix}
1\\0\\0\\0\end{pmatrix}},\quad\sigma_x\rightarrow{\begin{pmatrix}
0\\1\\0\\0\end{pmatrix}},\quad\sigma_y\rightarrow{\begin{pmatrix}
0\\0\\1\\0\end{pmatrix}},\quad\sigma_z\rightarrow{\begin{pmatrix}
0\\0\\0\\1\end{pmatrix}}.
\end{equation}
Due to the trivial eigenvalues $1$ corresponding to the identity operator, the maximum dimension of the non-trivial Jordan block is $3\times3$. We consider both $2\times2$ and $3\times3$ Jordan blocks and discuss them respectively.

\subsection{Transfer matrix with $2\times2$ Jordan Block}
We first consider the transfer matrix with $2\times2$ Jordan Block, which can be written as,
\begin{equation}\label{NDTFM2}  
\mathcal{M}_+^{(2)}={\begin{pmatrix}1&\textbf{0}_r\\
\textbf{0}_c&R^{(2)}\\
\end{pmatrix}},
\end{equation}
where $R^{(2)}={\begin{pmatrix}
r_1&0&l\\
0&r_2&0\\
0&0&r_1\\\end{pmatrix}}$, $\textbf{0}_r={\begin{pmatrix}0&0&0\end{pmatrix}}$, and $\textbf{0}_c=\textbf{0}_r^T$ with $T$ denoting matrix transposition.
Here, the parameters $r_1$, $r_2$, $l$ are related to the specific dual-unitary circuit, the $(2)$ in the upper right corner of $\mathcal{M}$ denotes a $2\times2$ Jordan Block in the transfer matrix $\mathcal{M}$. To realize a system with the above transfer matrix, one should find the corresponding local evolution operator under the dual-unitary conditions.
For simplicity, we consider the corresponding local dual-unitary evolution operator which satisfies the form,
\begin{equation}\label{}
U=(e^{i\Phi\sigma_y}\otimes e^{i\Phi\sigma_y})V[J](e^{i\phi\sigma_y}\otimes e^{i\phi\sigma_y}).
\end{equation}
Then, the transfer matrix can be written as,
\begin{equation}\label{ML1}
\begin{split}
\mathcal{M}_+^{(2)}(a)=&\frac{1}{2}\text{tr}_1[U^\dag(a \otimes \mathbbm{1}_2)U]\\
=&\frac{1}{2}\text{tr}_1[(\mathbbm{1}_2\otimes e^{-i\phi\sigma_y})V^\dag[J](e^{-i\Phi\sigma_y}ae^{i\Phi\sigma_y} \otimes \mathbbm{1}_2)\\
& V[J](\mathbbm{1}_2\otimes e^{i\phi\sigma_y})].\\
\end{split}
\end{equation}
On the other hand, the transfer matrix $\mathcal{M}_+^{(2)}$ corresponds to mapping as,
\begin{equation}\label{ML2}
\begin{split}
\mathcal{M}_+^{(2)}(\mathbbm{1}_2)=&\mathbbm{1}_2\\
\mathcal{M}_+^{(2)}(\sigma_x)=&r_1\sigma_x\\
\mathcal{M}_+^{(2)}(\sigma_y)=&r_2\sigma_y\\
\mathcal{M}_+^{(2)}(\sigma_z)=&r_1\sigma_z+l\sigma_x.\\
\end{split}
\end{equation}
The combinations of Eq~\eqref{ML1} and Eq~\eqref{ML2} give rise to,
\begin{equation}\label{SEQs2} 
	\begin{split}
		&\text{tr}_1[(\mathbbm{1}_2\otimes e^{-i\phi\sigma_y})V^\dag[J](e^{-i2\Phi\sigma_y}\sigma_x \otimes \mathbbm{1}_2)V[J](\mathbbm{1}_2\otimes e^{i\phi\sigma_y})]\\
		=&2r_1\sigma_x,\\
		&\text{tr}_1[(\mathbbm{1}_2\otimes e^{-i\phi\sigma_y})V^\dag[J](\sigma_y \otimes \mathbbm{1}_2)V[J](\mathbbm{1}_2\otimes e^{i\phi\sigma_y})]\\
		=&2r_2\sigma_y,\\
		&\text{tr}_1[(\mathbbm{1}_2\otimes e^{-i\phi\sigma_y})V^\dag[J](e^{-i2\Phi\sigma_y}\sigma_z \otimes \mathbbm{1}_2)V[J](\mathbbm{1}_2\otimes e^{i\phi\sigma_y})]\\
		=&2r_1\sigma_z+2l\sigma_x.\\
	\end{split}
\end{equation}
By solving Eq.~\eqref{SEQs2} and considering that the correlation function is not greater than 1, we can obtain
the constraint relations between the parameters $J$, $\Phi$, and $\phi$ can be obtained as,
\begin{equation}\label{CSTR2}
\begin{split}
&\Phi,\phi\neq\frac{n\pi}{4}\quad n\in\mathbbm{Z}\\
&|\cos{2\Phi}|>|\cos{2\phi}|\\
&\phi=\Phi-\frac{\pi}{4}\\
&J=\frac{1}{2}\arcsin{\left(\tan^2{2\Phi}\right)}.\\
\end{split}
\end{equation}
Thus, the parameters of the transfer matrix $\mathcal{M}_+^{(2)}$ can be calculated as,
\begin{equation}
\begin{split}
&r_1=\tan{2\Phi}\\
&r_2=r_1^2\\
&l=\tan^2{2\Phi}-1.\\
\end{split}
\end{equation}
As all other parameters can be determined by $\Phi$, therefore there is only one free parameter $\Phi$ in the transfer matrix. The $\Phi\neq\frac{n\pi}{4}$ thus gives a one-parameter family of dual-unitary quantum circuits whose transfer matrix has a second-order EPs.
Specifically,  $\Phi\in\left(\frac{n\pi}{2},\frac{n\pi}{2}+\frac{\pi}{8}\right]\bigcup\left[\frac{n\pi}{2}+\frac{3\pi}{8},\frac{n\pi}{2}+\frac{\pi}{2}\right)$ with $n\in\mathbbm{Z}$.

Now we can calculate the spatiotemporal correlation function at the light-cone edges as,
\begin{equation}\label{LCCF2}
\begin{split}
C^{\alpha,\alpha}&=r_1^{2t}, \quad \alpha=x,z\\
C^{y,y}&=r_2^{2t}\\
C^{x,z}&=2tlr_1^{2t-1}\\
C^{x,y}&=C^{y,z}=0.\\
\end{split}
\end{equation}
The results show that $C^{x,z}$ is a polynomial-enhanced exponential decay light-cone correlation function, while $C^{x,x}$ and $C^{z,z}$ are exponential decay. We investigate the light-cone correlation function numerically. As an example, we choose $\Phi=5\pi/48$, and then the other parameters can be calculated by the constraint relation Eq.~\eqref{CSTR2} as $\phi=7\pi/48$, and $J\approx0.3148$. Then the corresponding parameters of the transfer matrix Eq.~\eqref{NDTFM2} can be obtained as $r_1\approx0.7673$, $r_2\approx0.5888$, and $l\approx-0.4112$.

\begin{figure}[htbp] \centering
\includegraphics[width=8.9cm]{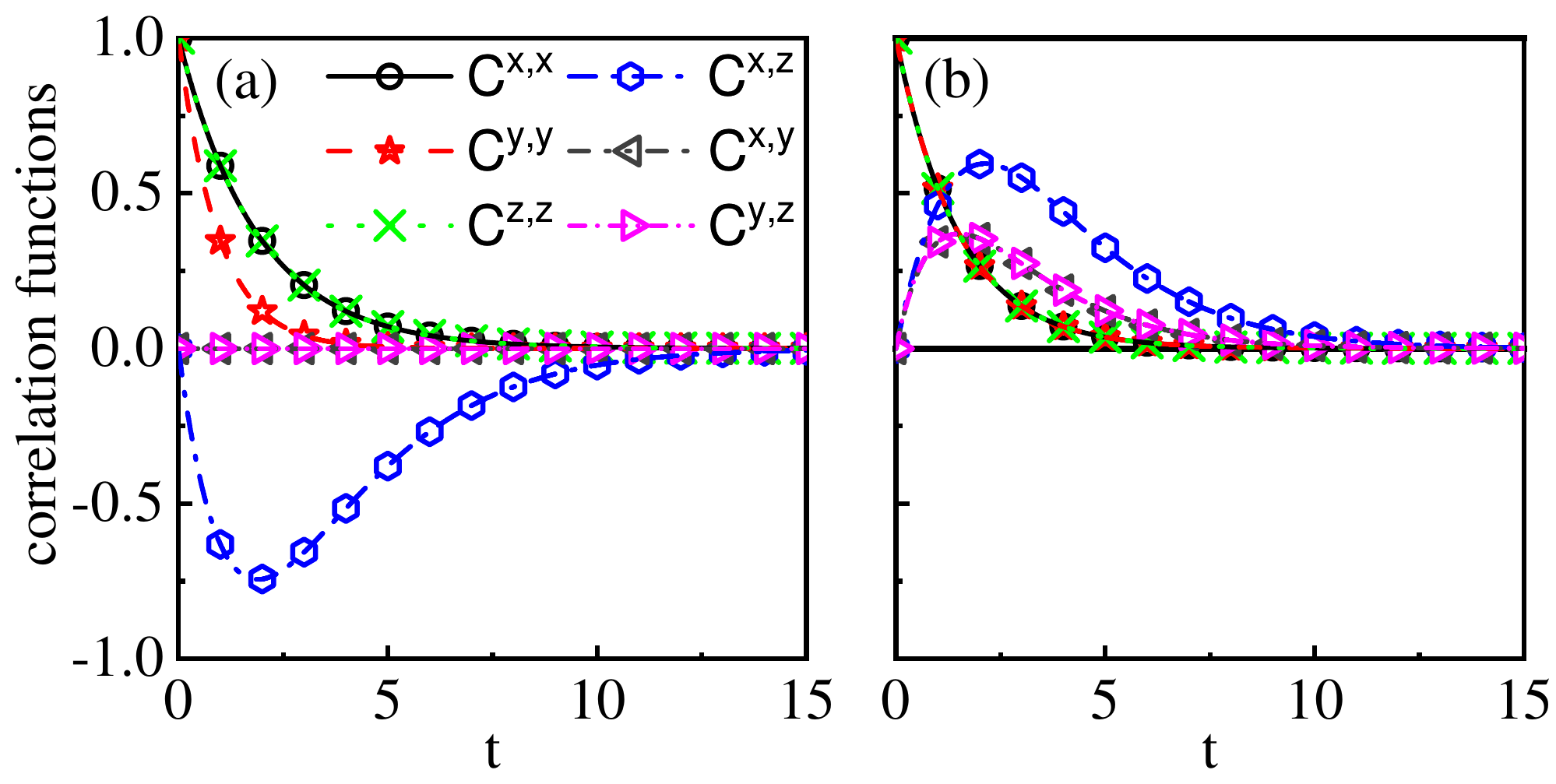}
\caption{(Color online). The light-cone correlation function for dual-unitary quantum circuits with EPs in the transfer matrix. (a) and (b) are the light-cone correlation functions for the transfer matrix with $2\times2$ and $3\times3$ Jordan block, respectively. The analytical results are plotted as lines, which are directly calculated by Eq.~\eqref{LCCF2} for (a) and Eq.~\eqref{LCCF3} for (b). The results denoted by symbols are calculated by the evolution of the dual-unitary quantum circuits.}\label{CRFfig}
\end{figure}

To observe the polynomial-enhanced light-cone correlation function, we calculate the light-cone correlation functions $C^{\alpha,\alpha} (\alpha=x,y,z)$, $C^{x,y}$, $C^{x,z}$, $C^{y,z}$
and plot in Fig.~\ref{CRFfig} (a) with lines. The results show the light-cone correlation function of $C^{x,z}$ exhibits the polynomial growth with an exponential decay. A remarkable feature is that the absolute value of the light-cone correlation functions may first arise to a maximum value at $t=\xi=-\frac{1}{2\log r}$ (the correlation length) and then decay to zero. In contrast, $C^{\alpha,\alpha} (\alpha=x,y,z)$ exhibits solely exponential decay with time, whose decay is faster than the polynomial-enhanced case. Moreover, $C^{x,y}$, and $C^{y,z}$ are zeros all the time because the Jordan block is $2\times2$ constructed by $x$ and $z$. This means the correlation between the operator $\sigma_y$ and operator $\sigma_x$ (or $\sigma_z$) is zero, as shown in Fig.~\ref{CRFfig} (a).

\subsection{Transfer matrix with $3\times3$ Jordan Block}
Furthermore, we consider the transfer matrix with the upper triangular formula, which reads,
\begin{equation}\label{NDTFM3}  
\mathcal{M}_+^{(3)}(a)={\begin{pmatrix}1&\textbf{0}_r\\
\textbf{0}_c&R^{(3)}\\
\end{pmatrix}},
\end{equation}
where $R^{(3)}=r\mathbbm{1}_3+l_1S_++l_2S_+^2$ with $S_+={\begin{pmatrix}0&1&0\\
0&0&1\\
0&0&0\\\end{pmatrix}}$ and $\mathbbm{1}_3$ is a 3-dimensional identity matrix.
Here, the parameters $r$, $l_1$, and $l_2$ are related to the specific dual-unitary circuit, and the $R^{(3)}$  can be decomposed into a standard $3\times3$ Jordan block. We keep the parameterization of $R^{(3)}$ instead of a standard Jordan block as it is more convenient for the following derivation. To realize a system with the above transfer matrix, one should find the corresponding local evolution operator under the dual-unitary conditions.
For simplicity, we consider the corresponding local evolution operator as,
\begin{equation}\label{}
U=(u_+\otimes u_-)V[J](v_+\otimes v_-),
\end{equation}
where $u_-=u_+=e^{i\Psi\sigma_z}e^{i\Phi\sigma_y}$ and $v_-=v_+=e^{i\phi\sigma_y}e^{i\varphi\sigma_z}$.
Then, the transfer matrix $\mathcal{M}_+^{(3)}$ can be constructed as,
\begin{equation}\label{ML1_3}
\begin{split}
\mathcal{M}_+^{(3)}=&\frac{1}{2}\text{tr}_1[U^\dag(a \otimes \mathbbm{1}_2)U]\\
=&\frac{1}{2}\text{tr}_1[(\mathbbm{1}_2\otimes v_-^\dag)V^\dag[J](u_+^\dag au_+ \otimes \mathbbm{1}_2)V[J](\mathbbm{1}_2\otimes v_-)].\\
\end{split}
\end{equation}
On the other hand, the transfer matrix $\mathcal{M}_+^{(3)}$ corresponds to a mapping as,
\begin{equation}\label{ML2_3}
\begin{split}
\mathcal{M}_+^{(3)}(\mathbbm{1}_2)=&\mathbbm{1}_2\\
\mathcal{M}_+^{(3)}(\sigma_x)=&r\sigma_x\\
\mathcal{M}_+^{(3)}(\sigma_y)=&r\sigma_y+l_1\sigma_x\\
\mathcal{M}_+^{(3)}(\sigma_z)=&r\sigma_z+l_2\sigma_x+l_1\sigma_y.\\
\end{split}
\end{equation}
By combining Eq~\eqref{ML1_3} and Eq~\eqref{ML2_3}, we have,
\begin{equation}\label{SEQs3} 
	\begin{split}
		&\text{tr}_1[(\mathbbm{1}_2\otimes v_-^\dag)V^\dag[J](u_+^\dag \sigma_xu_+ \otimes \mathbbm{1}_2)V[J](\mathbbm{1}_2\otimes v_-)]\\
		=&2r\sigma_x,\\
		&\text{tr}_1[(\mathbbm{1}_2\otimes v_-^\dag)V^\dag[J](u_+^\dag \sigma_yu_+ \otimes \mathbbm{1}_2)V[J](\mathbbm{1}_2\otimes v_-)]\\
		=&2(r\sigma_y+l_1\sigma_x),\\
		&\text{tr}_1[(\mathbbm{1}_2\otimes v_-^\dag)V^\dag[J](u_+^\dag \sigma_zu_+ \otimes \mathbbm{1}_2)V[J](\mathbbm{1}_2\otimes v_-)]\\
		=&2(r\sigma_z+l_1\sigma_y+l_2\sigma_x).\\
	\end{split}
\end{equation}
By solving Eq.~\eqref{SEQs3}, the parameters $J$, $\Psi$, $\Phi$, $\phi$, and $\varphi$ should satisfy constraint relation,
\begin{equation}\label{CSTR}
\begin{split}
&\Psi,\Phi,\phi,\varphi\neq\frac{n\pi}{4}\quad n\in\mathbbm{Z}\\
&|\cos{2\Phi}|>|\cos{2\phi}|\\
&\sin^2{2\phi}\cos{2\phi}=\sin^2{2\Phi}\cos{2\Phi}\\
&\frac{\sin{2\Phi}}{\sin{2\phi}}=-\tan{2\varphi}\\
&\Psi=\varphi-\frac{1}{4}\pi\\
&J=\frac{1}{2}\arcsin{\left(\tan^3{2\varphi}\right)}.
\end{split}
\end{equation}
Therefore, the key parameters of the transfer matrix $\mathcal{M}_+^{(3)}$ can be calculated as,
\begin{equation}
\begin{split}
r&=\tan^2{2\varphi}\\
l_1&=\tan{2\varphi}(1-\tan^2{2\varphi})\\
l_2&=1-\tan^2{2\varphi}.\\
\end{split}
\end{equation}
Now only one free parameter is left, and thus  $\varphi\neq\frac{n\pi}{4}$ give rise to a one-parameter family of the dual-unitary quantum circuit with a third-order EPs in the transfer matrix. Specifically, $\varphi \in\left(\frac{n\pi}{2},\frac{n\pi}{2}+\frac{\pi}{8}\right]\bigcup\left[\frac{n\pi}{2}+\frac{3\pi}{8},\frac{n\pi}{2}+\frac{\pi}{2}\right)$ with $n\in\mathbbm{Z}$.

Finally, we calculate the spatiotemporal correlation function at the light-cone edges as,
\begin{equation}\label{LCCF3}
\begin{split}
C^{\alpha,\alpha}&=r^{2t}, \quad \alpha=x,y,z\\
C^{x,z}&=2tl_2r^{2t-1}+\frac{2t(2t-1)}{2}l_1^2r^{2t-2}\\
C^{x,y}&=2tl_1r^{2t-1}\\
C^{y,z}&=2tl_1r^{2t-1}.\\
\end{split}
\end{equation}
The results show that the polynomial-enhanced light-cone correlation functions can be obtained as $C^{x,y}$, $C^{x,z}$, $C^{y,z}$ with an exponential decay.
To visualize the polynomial growth light-cone correlation function, we show the above results in Fig.~\ref{CRFfig}(b) with the specific parameters. Here, we choose $\Phi=2\pi/15$, and other parameters can be calculated by the constraint relation Eq.~\eqref{CSTR} as $\Psi\approx-0.4339$, $\phi\approx0.5348$, $\varphi\approx0.3515$, and $J\approx0.3270$. Then the corresponding parameters of the transfer matrix Eq.~\eqref{NDTFM3} can be given as $r\approx0.7180$, $l_1\approx-0.2390$, $l_2\approx0.2820$.
The results show that the light-cone correlation function of $C^{x,y}$, $C^{x,z}$, and $C^{y,z}$ exhibit the polynomial growth with an exponential decay. In contrast, $C^{\alpha,\alpha}$ $(\alpha=x,y,z)$ shows solely exponential decay.

\subsection{Kicked XXZ spin chain}

We further illustrate that the dual-unitary quantum circuit constructed in the above can be approximately mapped as a dynamical evolution of a Floquet system. In this regard, the polynomial-enhanced light-cone correlation function can also appear in the language of Hamiltonian evolution. Such a Floquet system can also be considered as a Trotter decomposition form of a XXZ model with two kick external magnetic fields. The corresponding Hamiltonian can be written as,
\begin{equation}\label{KXXZM}
H=H_{XXZ}+\sum_{m=1}^{\infty}\delta\left(t-\frac{m}{2}\right)(H_Z+H_Y),
\end{equation}
where $\delta(t-\frac{m}{2})$ is Dirac delta function and,
\begin{equation}\label{SXXZ}
\begin{split}
H_{XXZ}=&H_{o}+H_{e}\\
H_Z=&H_{1}+H_{4}\\
H_Y=&H_{3}+H_{2},\\
\end{split}
\end{equation}
with,
\begin{equation}
\begin{split}
H_o=&\sum_{n\in odd}^N\frac{\pi}{4}\sigma_x^n\sigma_x^{n+1}+\frac{\pi}{4}\sigma_y^n\sigma_y^{n+1}+J\sigma_z^n\sigma_z^{n+1}\\
H_e=&\sum_{n\in even}^N\frac{\pi}{4}\sigma_x^n\sigma_x^{n+1}+\frac{\pi}{4}\sigma_y^n\sigma_y^{n+1}+J\sigma_z^n\sigma_z^{n+1}\\
H_{1}=&\sum_{n=1}^N -\varphi\sigma_z^n,~~~
H_{2}=\sum_{n=1}^N -\phi\sigma_y^n,\\
H_{3}=&\sum_{n=1}^N -\Phi\sigma_y^n,~~~
H_{4}=\sum_{n=1}^N -\Psi\sigma_z^n.\\
\end{split}
\end{equation}
Then, by modulating the parameters during the trotter decomposition, one can obtain a dual-unitary quantum circuit form, which can be realized as the evolution operator of the Floquet systems. This Floquet system consists of a set of unitary evolution operators,
\begin{equation}\label{}
\begin{split}
U_\eta=&e^{-iH_{\eta}},\\
\end{split}
\end{equation}
where $\eta \in \{1,2,3,4,o,e\}$.

\begin{figure}[htbp] \centering
\includegraphics[width=8.5cm]{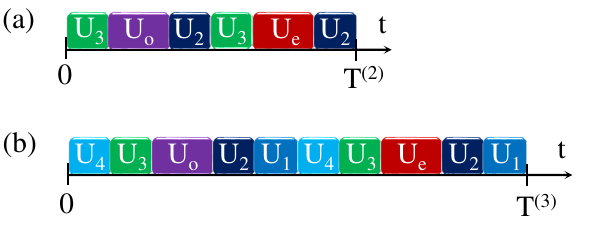}
\caption{(Color online). The decomposition of evolution operator for the Floquet system of XXZ spin chain. (a) is the Floquet system for transfer matrix with $2\times2$ Jordan block in a period $T^{(2)}$. (b) is the Floquet system for transfer matrix with $3\times3$ Jordan block in a period $T^{(3)}$. }\label{FQSfig}
\end{figure}

For transfer matrix with $2\times2$ Jordan Block $\mathcal{M}^{(2)}_+$, the evolution operator in a period reads,
\begin{equation}\label{EVO2}
U_T^{(2)}=U_2U_eU_3U_2U_oU_3.
\end{equation}
The Fig.~\ref{FQSfig}(a) shows the scheme of the evolution operator for $\mathcal{M}^{(2)}_+$.

The corresponding evolution operator for $\mathcal{M}^{(3)}_+$ in a period can be read,
\begin{equation}\label{EVO3}
U_T^{(3)}=U_1U_2U_eU_3U_4U_1U_2U_oU_3U_4.
\end{equation}
The Fig.~\ref{FQSfig}(b) shows the decomposition of the evolution operator for $\mathcal{M}^{(3)}_+$.

\begin{figure}[htbp] \centering
\includegraphics[width=8.5cm]{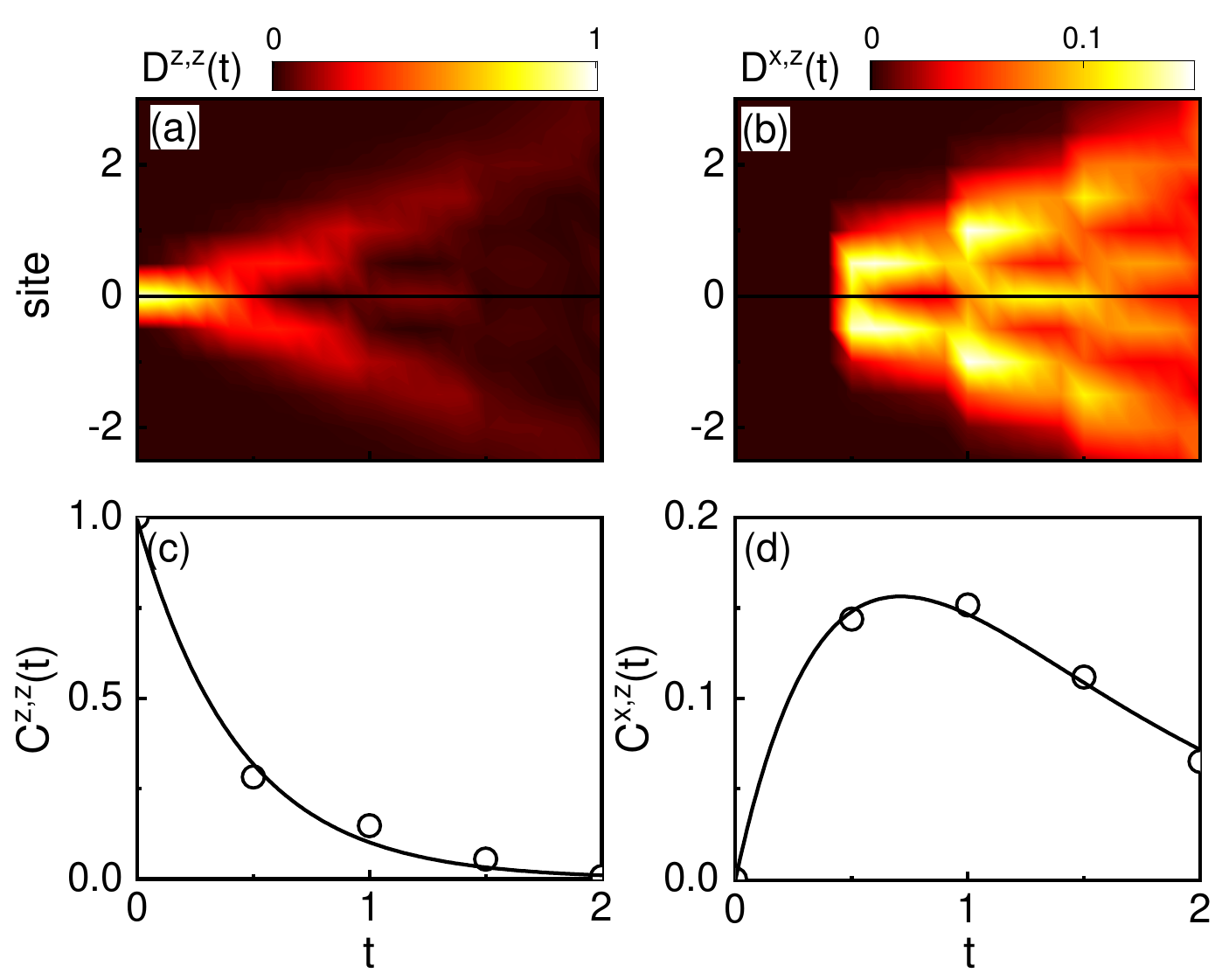}
\caption{(Color online). Spatiotemporal correlation function for the Kicked XXZ spin chain corresponding to $2\times2$ Jordan block. (a) and (b) are the spatiotemporal correlation function. (c) and (d) are the corresponding light-cone correlation function (symbol) of the spatiotemporal correlation function (a) and (b), respectively. The lines are fitting curves $ab^{-ct}$ for $C^{z,z}(t)$ and $atb^{-ct}$ for $C^{x,z}(t)$, respectively. }\label{DC2fig}
\end{figure}

\begin{figure}[htbp] \centering
\includegraphics[width=8.5cm]{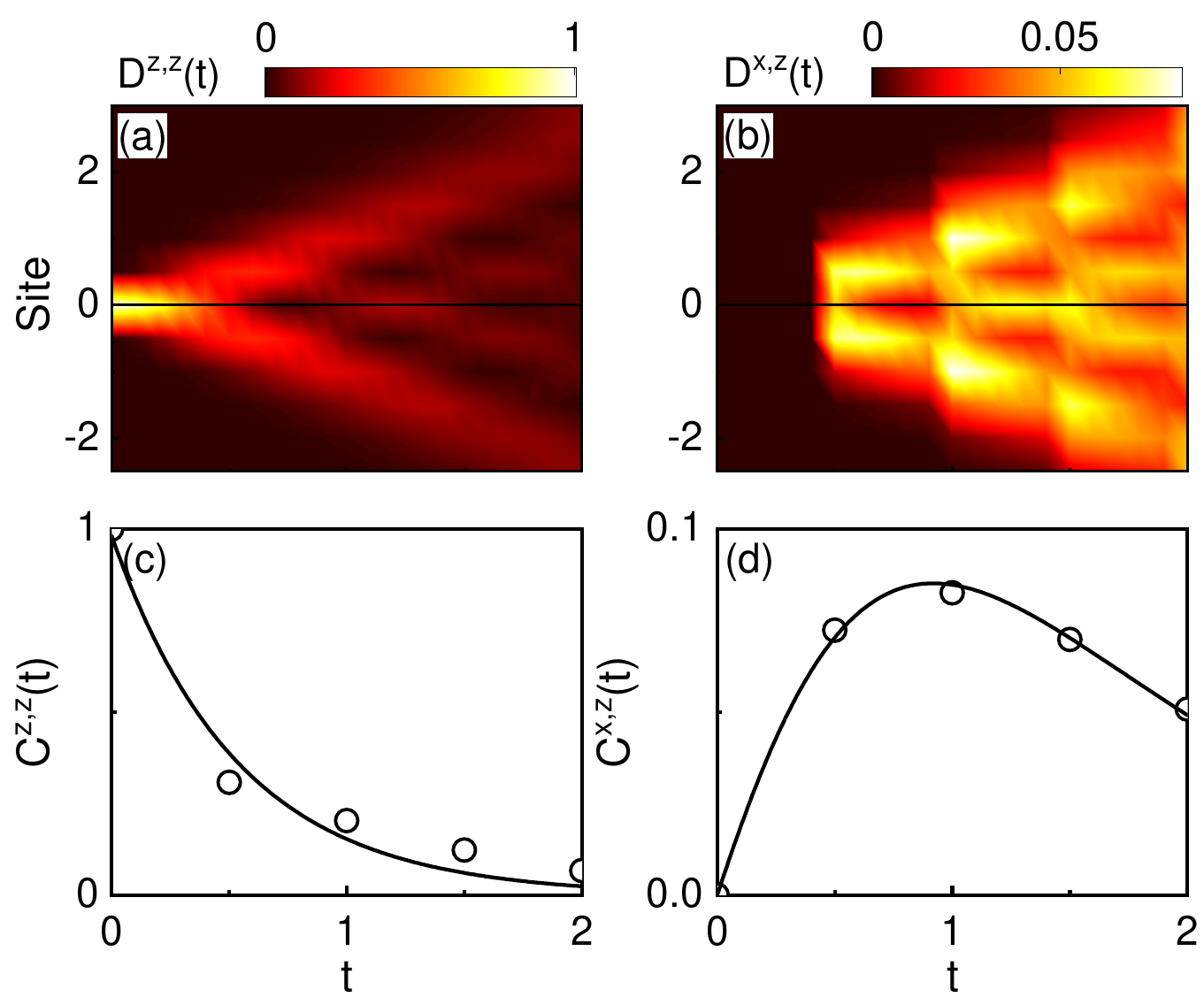}
\caption{(Color online). Spatiotemporal correlation function for the Kicked XXZ spin chain corresponding to $3\times3$ Jordan block. (a) and (b) are the spatiotemporal correlation function. (c) and (d) are the corresponding light-cone correlation function (symbol) of the spatiotemporal correlation function (a) and (b), respectively. The lines are fitting curves $ab^{-ct}$ for $C^{z,z}(t)$ and $(a_1t+a_2t^2)b^{-ct}$ for $C^{x,z}(t)$, respectively.}\label{DC3fig}
\end{figure}
Furthermore, we calculate the dynamical correlation function of an infinite temperature state under the kick XXZ model Eq.~\eqref{KXXZM}. Interestingly, the light-cone correlation function above can be fitted by the function type of corresponding analytical results Eq.~\eqref{LCCF2} and Eq.~\eqref{LCCF3}, as shown in Fig.~\ref{DC2fig} and Fig.~\ref{DC3fig}. Moreover, the dynamical correlation function is mainly concentrated on the light-cone and the light-cone correlation function can be fitting the polynomial modified exponential decay, which implies that this model can be an approximation of the dual-unitary circuits.


\section{properties near EPs}
\label{s4}

In this section, we first give the analytical results of the correlation function approaching EPs. Then, we analyze behaviors of correlation functions by discrete Fourier transformation and $Z$ transformation. It is shown that $Z$ transformation can well capture the polynomial modification of the correlation function and thus can distinguish the behaviors approaching and at EPs of the transfer matrix.

\subsection{$2\times2$ Jordan Block}
\begin{figure}[htbp] \centering
\includegraphics[width=8.5cm]{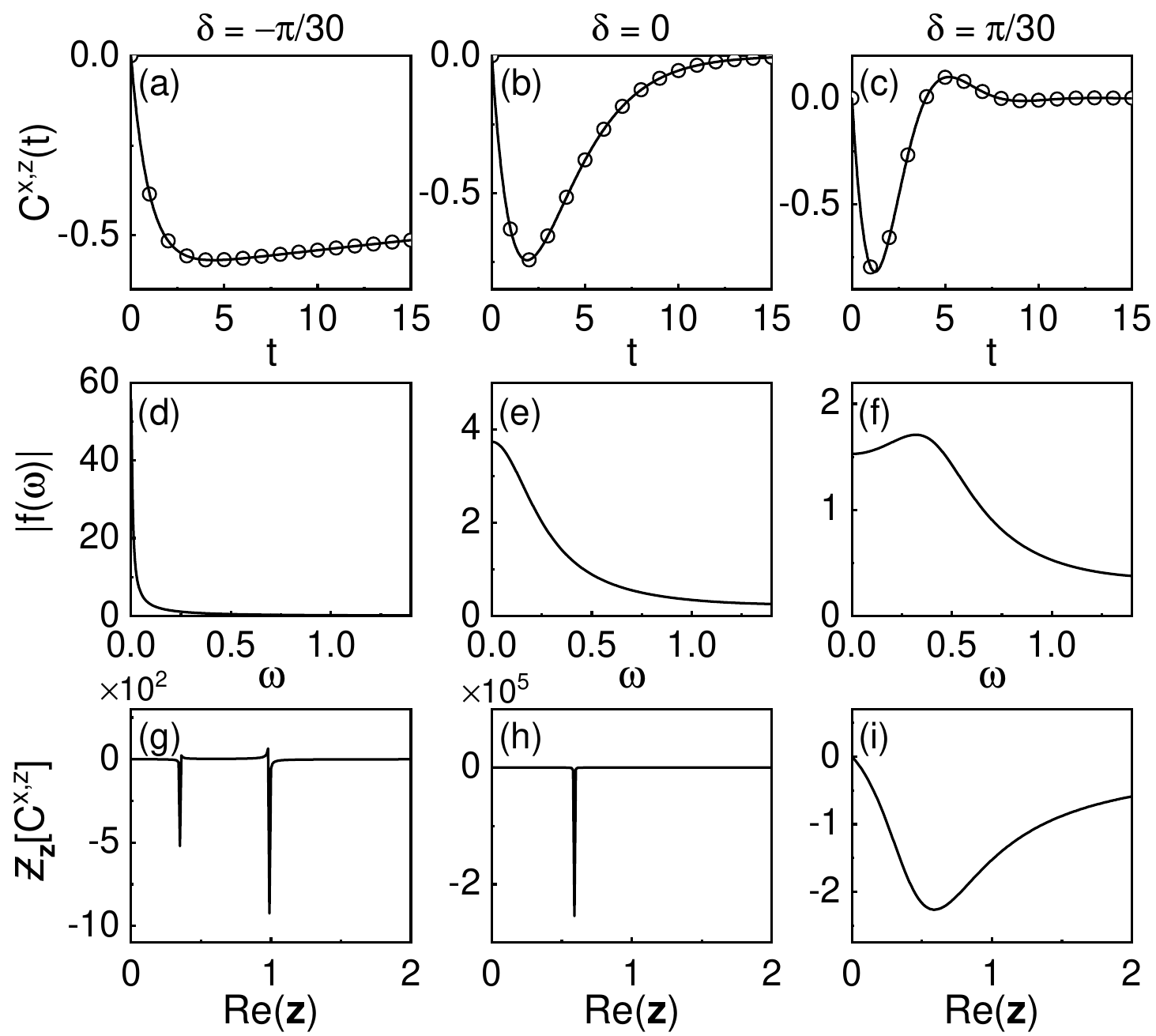}
\caption{The correlation function, Fourier amplitude, and $Z$ transformation of correlation function $C^{x,z}_{\delta}$ for transfer matrix with $2\times2$ Jordan block case. (a), (b), and (c) is the correlation function $C^{x,z}_{\delta}$ with different parameters $\delta$. (d), (e), and (f) is the corresponding Fourier amplitude of correlation function $C^{x,z}_{\delta}$. (g), (h), and (i) is the corresponding $Z$ transformation of the correlation function $C^{x,z}_{\delta}$.   }\label{EP2fig}
\end{figure}

For $2\times2$ Jordan Block case, the form of the transfer matrix away from EPs can be written as,
\begin{equation}\label{TSM2delta}
\begin{split}
\mathcal{M}_+^{(2)}=&{\begin{pmatrix}1&0&0&0\\0&\tan{2\Phi}\cos{2\delta}&0&\frac{-\cos{(4\Phi-2\delta)}+\frac{1}{2}\sin{2\delta}\sin{4\Phi}}{\cos^2{2\Phi}}\\
0&0&\sin{2J}&0\\0&\tan{2\Phi}\sin{2\delta}&0&\frac{\sin{(4\Phi+2\delta)}-\frac{1}{2}\cos{2\delta}\sin{4\Phi}}{\cos^2{2\Phi}}\end{pmatrix}},\\
\end{split}
\end{equation}
where $\delta$ is the parameter that causes the matrix to deviate from EPs. Accordingly, the constraint relations between the parameters $J$, $\Phi$, and $\phi$ are rewritten as,
\begin{equation}\label{CSTR2delta}
\begin{split}
&\Phi,\phi\neq\frac{n\pi}{4}\quad n\in\mathbbm{Z}\\
&|\cos{2\Phi}|>|\cos{2\phi}|\\
&\phi=\Phi-\frac{\pi}{4}-\delta\\
&J=\frac{1}{2}\arcsin{\left(\tan^2{2\Phi}\right)}.\\
\end{split}
\end{equation}
When $\delta\neq0$, the eigenvalues of the transfer matrix Eq.~\eqref{TSM2delta} are $E_1=\frac{\sin{(4\Phi-2\delta)}}{2\cos^2{2\Phi}}+\frac{\Delta}{2}$, $E_2=\frac{\sin{(4\Phi-2\delta)}}{2\cos^2{2\Phi}}-\frac{\Delta}{2}$, $E_3=\sin{2J}$, and $E_4=1$ with $\Delta=\sqrt{-\frac{\sin{2\delta}\sin{(8\Phi-2\delta)}}{\cos^4{2\Phi}}}$. Due to the transfer matrix Eq.~\eqref{TSM2delta} can be diagonalized, the polynomial correlation function $C^{x,z}$ turn into the form,
\begin{equation}
\begin{split}
C^{x,z}_{\delta}(t)=&\frac{l'}{\Delta}\left[\left(E_1\right)^{2t}-\left(E_2\right)^{2t}\right],\\
\end{split}
\end{equation}
where $l'=\frac{-\cos{(4\Phi-2\delta)}+\frac{1}{2}\sin{2\delta}\sin{4\Phi}}{\cos^2{2\Phi}}$. This result exhibits exponential decay rather than polynomial behavior. Moreover, when $\delta<0$, $\Delta$ is a real number. For this case, the correlation function exhibits the superposition of two different modes of exponential decay, which can be observed by $Z$ transformation as,
\begin{equation}\label{}
\mathcal{Z}_\mathbf{z}[C^{x,z}_{\delta\neq0}(t)]=\frac{l'\mathbf{z}}{\Delta}\left[\frac{1}{\mathbf{z}-{E_1}^2}-\frac{1}{\mathbf{z}-{E_2}^2}\right].\\
\end{equation}
For $\delta=0$, the exceptional points, the $Z$ transformation is,
\begin{equation}\label{}
\mathcal{Z}_\mathbf{z}[C^{x,z}(t)]=\frac{2lr_1\mathbf{z}}{(\mathbf{z}-{r_1}^2)^2},
\end{equation}
which shows exponential decay with one mode. For $\delta>0$, $E_1$ and $E_2$ turn to complex numbers. Although the $Z$ transformation for $C^{x,z}_{\delta>0}$ is the same as the case of $\delta<0$, the result of $\mathcal{Z}_\mathbf{z}[C^{x,z}_{\delta>0}]$ is very different from the case of $\delta\leq0$, as shown in Fig.~\ref{EP2fig} (g) and (i).
Furthermore, due to $E_1$ and $E_2$ are complex numbers, the correlation function $C^{x,z}_{\delta>0}$ will show oscillation behavior. By introducing discrete Fourier transformation, we obtain the frequency function,
\begin{equation}
f_{\delta\neq0}(\omega)=\frac{l'}{\Delta}\left[\frac{1}{1-{E_1}^2e^{i2\omega}}-\frac{1}{1-{E_2}^2e^{i2\omega}}\right].
\end{equation}
By the way, we calculate the discrete Fourier transformation of the correlation function $C^{x,z}$ as,
\begin{equation}
\begin{split}
f(\omega)=\frac{2lr_1e^{-i2\omega}}{(e^{-i2\omega}-{r_1}^2)^2},
\end{split}
\end{equation}
for $\delta=0$.

As shown in Fig.~\ref{EP2fig}(a), (b), and (c), the properties of the correlation function $C^{x,z}$ are different for approaching and at EPs. For $\delta<0$, the correlation function $C^{x,z}$ shows exponential behavior with two modes, which include one mode decay and one mode growth. These two modes can be clearly depicted by $Z$ transformation with two points of divergence, as shown in Fig.~\ref{EP2fig} (g). Moreover, this behavior is fundamentally different from the case of $\delta=0$, which exhibits a polynomial enhancement with one mode exponential decay. Although Fig.~\ref{EP2fig} (a) and (b) show similar behavior, increasing firstly and then decreasing, the $Z$ transformation of case $\delta=0$ have only one point of divergence, as shown in Fig.~\ref{EP2fig} (h). However, for $\delta>0$, the correlation function $C^{x,z}$ shows oscillation with exponential decay [as shown in Fig.~\ref{EP2fig} (c)], which is significantly different from the previous two cases. On the other hand, as shown in Fig.~\ref{EP2fig} (d), (e) and, (f),
discrete Fourier analysis gives signals that are not that clear compared to $Z$ transformation.


As a consequence, the rich properties of the correlation functions $C^{x,z}$ for approaching and at EPs can be clearly observed by $Z$ transformation, which is due to the differences of the transfer matrices. For $\delta=0$, the coalesce of eigenvectors results in the correlation showing polynomial behavior. For $\delta<0$, the correlation shows two modes of exponential decay, because the transfer matrix can be diagonalized as a real spectrum. While, for $\delta>0$, due to the complex spectrum of the transfer matrix, the correlation function exhibits the oscillation behavior with one mode of exponential decay.

\subsection{$3\times3$ Jordan Block}
For $3\times3$ Jordan Block case, the form of the transfer matrix away from EPs is written as,
\begin{equation}\label{TSM3delta}
\begin{split}
\mathcal{M}_+^{(3)}=&{\begin{pmatrix}1&0&0&0\\
                              0&r'_1&l_3&l_2\\
                              0&l_4&r'_2&l_1\\
                              0&0&0&r\\ \end{pmatrix}}\\
\end{split}
\end{equation}
with parameters,
\begin{equation}\label{}
\begin{split}
r'_1=&\tan^2{2\varphi}\cos{2\delta}-\tan{2\varphi}(1-\tan^2{2\varphi})\sin{2\delta}\\
r'_2=&\tan^2{2\varphi}\cos{2\delta}\\
r=&\tan^2{2\varphi}\\
l_1=&\tan{2\varphi}(1-\tan^2{2\varphi})\\
l_2=&1-\tan^2{2\varphi}\\
l_3=&\tan{2\varphi}(1-\tan^2{2\varphi})\cos{2\delta}+\tan^2{2\varphi}\sin{2\delta}\\
l_4=&-\tan^2{2\varphi}\sin{2\delta},\\
\end{split}
\end{equation}
where $\delta$ is the parameter that causes the matrix to deviate from EPs. To simplify, we keep some constraint relations between the parameters $J$, $\Psi$, $\Phi$, $\phi$, and $\varphi$. Then we get the constraint relations as follows,
\begin{equation}\label{CSTR3delta}
\begin{split}
&\Psi,\Phi,\phi,\varphi\neq\frac{n\pi}{4}\quad n\in\mathbbm{Z}\\
&|\cos{2\Phi}|>|\cos{2\phi}|\\
&\sin^2{2\phi}\cos{2\phi}=\sin^2{2\Phi}\cos{2\Phi}\\
&\frac{\sin{2\Phi}}{\sin{2\phi}}=-\tan{2\varphi}\\
&\Psi=\varphi-\frac{1}{4}\pi-\delta\\
&J=\frac{1}{2}\arcsin{\left(\tan^3{2\varphi}\right)},
\end{split}
\end{equation}
which imply that only $\Psi$ changes $\delta$.
When $\delta\neq0$, the transfer matrix Eq.~\eqref{TSM3delta} can be diagonalized. The corresponding eigenvalues are $E_1=\frac{r'_1+r'_2+\Delta'}{2}$, $E_2=\frac{r'_1+r'_2-\Delta'}{2}$, $E_3=r$, and $E_4=1$ with $\Delta'=\sqrt{4l_3l_4+(r'_1-r'_2)^2}$. Then, the correlation function is written as,
\begin{equation}
\begin{split}
C^{x,z}_\delta(t)=&A_1E_1^{2t}+A_2E_2^{2t}+A_3E_3^{2t},
\end{split}
\end{equation}
where,
\begin{equation}\label{}
\begin{split}
A_1=&\frac{2l_3[l_2l_4+l_1(r-r'_1)]+(r'_1-r'_2+\Delta')[l_1l_3+l_2(r-r'_2)]}{2\Delta'[l_3l_4+(r'_1-r)(r-r'_2)]}\\
A_2=&\frac{2l_3[l_2l_4+l_1(r-r'_1)]+(r'_1-r'_2-\Delta')[l_1l_3+l_2(r-r'_2)]}{-2\Delta'[l_3l_4+(r'_1-r)(r-r'_2)]}\\
A_3=&-\frac{l_1l_3+l_2(r-r'_2)}{l_3l_4+(r'_1-r)(r-r'_2)}.\\
\end{split}
\end{equation}
These results imply that the correlation function will exhibit multiple modes of exponential decay rather than polynomial behavior. For $\delta<0$, $\Delta'$ is a real number. In this case, the correlation function $C^{x,z}$ is a superposition of three different exponential decay modes, which can be described by $Z$ transformation as,
\begin{equation}\label{}
\mathcal{Z}_\mathbf{z}[C^{x,z}_{\delta\neq0}(t)]=\mathbf{z}\left(\frac{A_1}{\mathbf{z}-{E_1}^2}+\frac{A_2}{\mathbf{z}-{E_2}^2}+\frac{A_3}{\mathbf{z}-{E_3}^2}\right).\\
\end{equation}
For $\delta=0$, the $Z$ transformation of correlation function $C^{x,z}$ is,
\begin{equation}\label{}
\mathcal{Z}_\mathbf{z}[C^{x,z}(t)]=\frac{2 l_2 r \mathbf{z}}{(\mathbf{z}-r^2)^2}+\frac{{l_1}^2 (3r^2+\mathbf{z})}{(\mathbf{z}-r^2)^3},
\end{equation}
which shows that the correlation function exhibits exponential decay with one mode. However, for $\delta>0$, $\Delta'$ is an image number, which implies that the correlation function will exhibit oscillation behavior. By introducing the discrete Fourier transformation, we obtain the function of frequency as,
\begin{equation}\label{}
f_{\delta\neq0}(\omega)=\frac{A_1}{1-{E_1}^2e^{i2\omega}}+\frac{A_2}{1-{E_2}^2e^{i2\omega}}+\frac{A_3}{1-{E_3}^2e^{i2\omega}}.
\end{equation}
Furthermore, we calculate the discrete Fourier transformation of the correlation function $C^{x,z}$ as,
\begin{equation}
f(\omega)=\frac{2 l_2 r e^{-i2\omega}}{(e^{-i2\omega}-r^2)^2}+\frac{{l_1}^2 (3r^2+e^{-i2\omega})}{(e^{-i2\omega}-r^2)^3}
\end{equation}
for $\delta=0$.

\begin{figure}[htbp] \centering
\includegraphics[width=8.5cm]{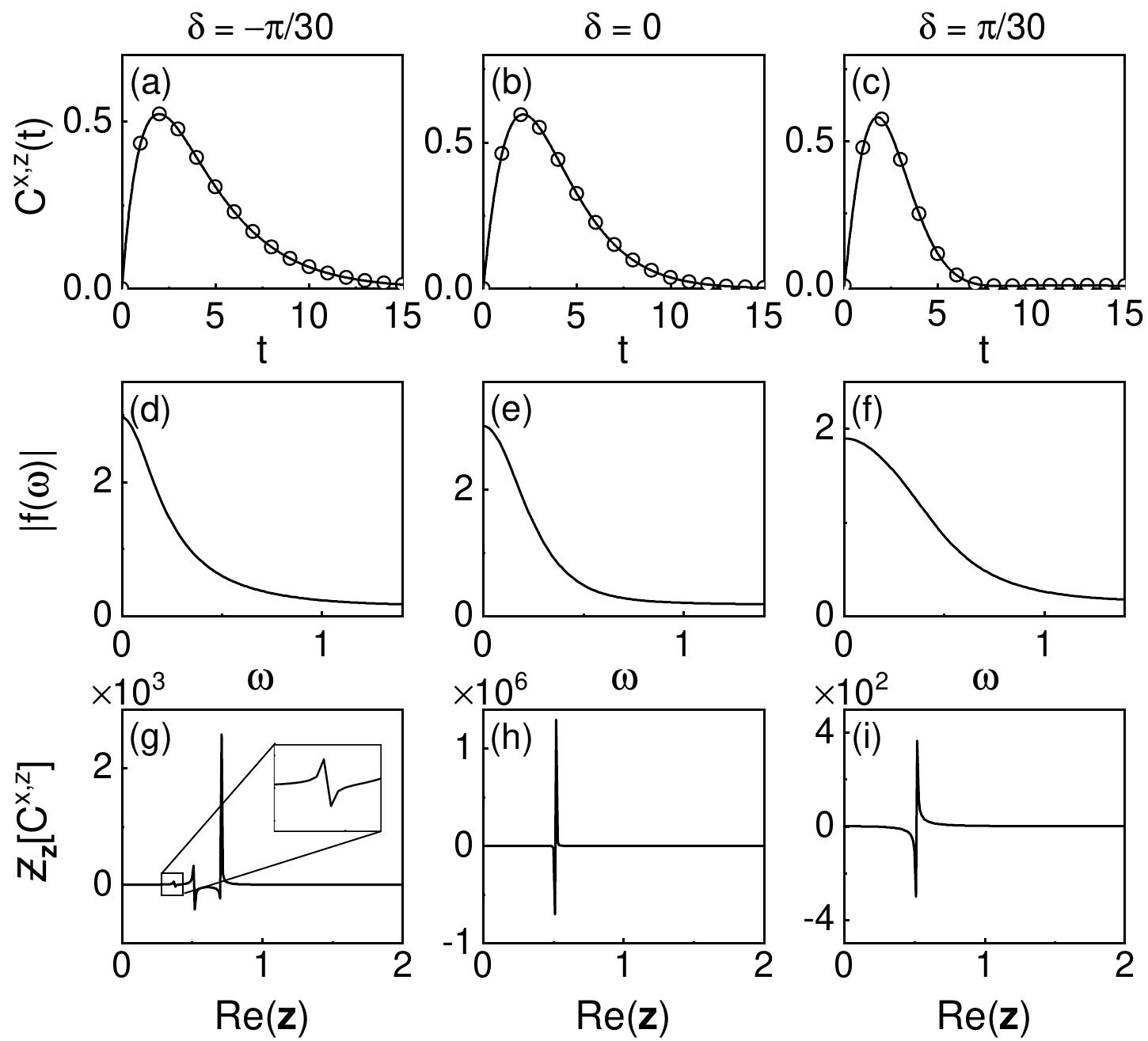}
\caption{The correlation function, Fourier amplitude, and $Z$ transformation of correlation function $C^{x,z}_{\delta}$ for transfer matrix with $3\times3$ case. (a), (b), and (c) is the correlation function $C^{x,z}_{\delta}$ with different parameters $\delta$. (d), (e), and (f) is the corresponding Fourier amplitude of correlation function $C^{x,z}_{\delta}$. (g), (h), and (i) is the corresponding $Z$ transformation of the correlation function $C^{x,z}_{\delta}$. }\label{EP3fig}
\end{figure}

Although the correlation functions look very similar [shown in Fig.~\ref{EP3fig} (a), (b), and (c)], their dynamical properties are fundamentally different. Specifically, for $\delta<0$, the correlation function shows exponential decay behavior with three modes, which is clearly described by $Z$ transformation with three divergent points, as shown in Fig.~\ref{EP3fig} (g). However, for $\delta>0$, the $Z$ transformation shows only one divergent point, because two eigenvalues of the transfer matrix turn into complex numbers, which will make the correlation function show oscillating rather than exponential decay behaviors. Moreover, this is different from the case of $\delta=0$, the polynomial-enhanced with exponential decay, whose $Z$ transformation appears as a higher order divergence point, as shown in Fig.~\ref{EP3fig} (h) and (i). In contrast, we find that the discrete Fourier transformation can not clearly distinguish the different correlation functions with different dynamical properties for approaching and at EPs.

Overall, the dynamical properties of the correlation function for $3\times3$ Jordan block case are similar to the $2\times2$ Jordan block case. While, for $3\times3$ Jordan block case, the correlation function will add a mode when their transfer matrix approaches EPs. Correspondingly, when the transfer matrix is located at EPs, the polynomial enhancement of correlation function $C^{x,z}$ corresponding to $3\times3$ Jordan block will be stronger than that of $2\times2$ Jordan block [see Eq.~\eqref{LCCF2} and Eq.~\eqref{LCCF3}].

\section{conclusion}
\label{s5}
In summary, we have constructed the local evolution gates, whose transfer matrix includes $2\times2$ and $3\times3$ Jordan blocks for $d=2$ case. By solving the relationship between the transfer matrix and spatiotemporal correlation function, we have derived the constrained relationship of the parameters of the local evolution operator to the transfer matrix with $2\times2$ and $3\times3$ Jordan blocks for $d=2$ case. Moreover, we have proposed the Hamiltonian evolution of a kick XXZ mode as an approximation
of the dual-unitary circuits, due to both of them having similar behaviors of the spatiotemporal correlation functions.
Lastly, we have shown that distinct behaviors of correlation functions approaching and at EPs can be obtained by $Z$ transformation.

Furthermore,  we point out that the construction may be generalized to $d>2$ cases. Firstly, by constructing a $d^2\times d^2$ nondiagonal transfer matrix in a dual-unitary circuit with $d>2$, one can observe the polynomial modified behaviors of the spatiotemporal correlation function. Then, by solving the relationship between the transfer matrix and spatiotemporal correlation function, one can derive the constrained relationship of the parameters of the local evolution operator for $d>2$ cases. Finally, one can use such parameters to build a Floquet system, which can realize the corresponding dual-unitary circuit approximately.

\bibliographystyle{apsrev4-1}
\begin{acknowledgments}
 This work was supported by the National Natural Science Foundation of China (Grant No.12375013) and the Guangdong Basic and Applied Basic Research Fund (Grant No.2023A1515011460).
\end{acknowledgments}
\normalem

\bibliography{Ref}
\end{document}